\documentclass[english,twocolumn,aps,prb,superscriptaddress,]{revtex4-2}
\usepackage{babel}
\usepackage{amsmath,mathtools}
\usepackage{amssymb}
\usepackage{graphicx}
\usepackage{tabularx}
\usepackage{lipsum}
\usepackage{xcolor}
\begin{document}
\title{Dynamic Manipulation of Non-Hermitian Skin Effect through Frequency in Topolectrical Circuits}
\author{S M Rafi-Ul-Islam }
\email{e0021595@u.nus.edu}
\selectlanguage{english}%
\affiliation{Department of Electrical and Computer Engineering, National University of Singapore, Singapore}
\author{Zhuo Bin Siu}
\email{elesiuz@nus.edu.sg}
\selectlanguage{english}%
\affiliation{Department of Electrical and Computer Engineering, National University of Singapore, Singapore}
%\author{Haydar Sahin}
%\email{sahinhaydar@u.nus.edu}
%\selectlanguage{english}%
%\affiliation{Department of Electrical and Computer Engineering, National University of Singapore, Singapore}
%\affiliation{Institute of High Performance Computing, A*STAR, Singapore}
\author{Md. Saddam Hossain Razo}
\email{shrazo@u.nus.edu}
\affiliation{Department of Electrical and Computer Engineering, National University of Singapore, Singapore 117583, Republic of Singapore}
\author{Mansoor B.A. Jalil}
\email{elembaj@nus.edu.sg}
\selectlanguage{english}%
\affiliation{Department of Electrical and Computer Engineering, National University of Singapore, Singapore}
\begin{abstract}
One of the most fascinating phenomena in non-Hermitian systems is the extensive accumulation of the bulk eigenstates under open-boundary conditions which is known as the non-Hermitian skin effect (NSHE). Here, we propose a switchable NHSE in a topolectrical (TE) set-up which can be turned on or off simply by varying the driving frequency without any modification to the physical circuit. Specifically,  we consider a coupled system consisting of two non-Hermitian Hatano-Nelson chains where each node of one chain is connected to neighboring node of the other chain via resistive couplings with opposite signs for the two coupling directions. Interestingly, the NHSE is switched on only if the driving frequency is greater than a certain critical frequency. Conversely, the NHSE in the coupled system is switched off when the frequency falls below the critical value, even though the individual uncoupled chains still exhibit the NHSE. This frequency-controlled NHSE may pave the way for many possible applications including non-Hermitian sensors where the driving frequency can manipulate the current and voltage localization. 
\end{abstract}
\maketitle
\section{Introduction} 
Owing to the plethora of exotic and unconventional phenomena \cite{rafi2021non,PhysRevResearch.4.043108,gao2020anomalous,rafi2022interfacial,lu2021magnetic,PhysRevB.110.045444,xu2022unconventional,rafi2024twisted,fan2022emerging} displayed by non-Hermitian systems, which have no similar counterparts in Hermitian systems, such systems have attracted much attention in recent years \cite{gong2018topological,bergholtz2021exceptional,leykam2017edge}. One characteristic feature in non-Hermitian systems is the exponential accumulation of bulk states at the boundaries under open boundary conditions (OBC), which is known as the non-Hermitian skin effect (NHSE) \cite{rafi2021topological,siu2023terminal,rafi2024saturation,longhi2019probing,rafi2022type,song2019non,rafi2022unconventional,longhi2020unraveling}. In general, the properties of non-Hermitian systems are highly  sensitive to the choice of boundary conditions \cite{lee2019anatomy,guo2021exact,rafi2022system,edvardsson2022sensitivity}. For example, the bulk eigenenergy spectra under OBC may have a very different distribution on the complex energy plane compared to the periodic boundary condition (PBC) spectra. This dichotomy constitutes the collapse of the conventional bulk-boundary correspondence (BBC) \cite{xiong2018does,jin2019bulk,xiao2020non}. As an archetypal example of a non-Hermitian system with non-reciprocal coupling, a Hatano-Nelson chain exhibits accumulation of its bulk states  at one edge of the chain and its corresponding OBC spectrum lies on the real energy axis regardless of the size of the open chain \cite{schindler2021dislocation,roccati2021non}.

An added attraction of research into non-Hermitian systems is their ready implementation under various platforms such as photonic systems \cite{silveirinha2019topological,pan2018photonic}, ultra-cold atoms \cite{zhou2021engineering,ruschhaupt2006improvement}, metamaterials \cite{ghatak2020observation,hou2020topological}, optical systems \cite{jin2017topological,longhi2021non}, ring resonators \cite{ao2020topological,lin2021square,gao2022non} and topolectrical (TE) circuits \cite{rafi2021non,zhang2023anomalous,rafi2022type,zou2021observation,rafi2024chiral,rafi2020realization,sahin2023impedance,rafi2021topological,helbig2020generalized,rafi2023valley,zhang2022complex,hofmann2020reciprocal,shang2022experimental,liu2020gain}. Of all these platforms, TE circuits based on basic circuit elements have the especial advantages of flexibility, accessibility, and relative freedom in setting the coupling connections between lattice and sublattice sites \cite{kotwal2021active,rafi2023valley,lee2018topolectrical,rafi2020topoelectrical,dong2021topolectric,rafi2020anti,zhang2023anomalous,rafi2020realization,zhang2020topolectrical,kotwal2021active,lee2018topolectrical}. However, despite the relative ease of implementation of non-Hermiticity in TE circuits, there may be stability issues related to the use of active elements such as operational amplifiers which form part of a component called impedance converters with current inversion (INICs) \cite{helbig2020generalized,zou2021observation}. INICs are crucial in realizing non-reciprocal coupling, which is required to confer non-Hermiticity to TE circuits, as well as tune and control the degree of non-Hermiticity in the circuits. Hence, the availability of an alternative means to tune the non-Hermiticity without requiring different circuit set-ups would greatly ease the experimental implementation of non-Hermitian phenomena in TE circuits. 

 In this letter, we propose an alternative approach to modulating the NHSE phenomenon in a TE circuit implementation of a non-Hermitian system by exploiting the frequency dependence of the admittance values of inductors and capacitors. We constructed a TE circuit consisting of a pair of coupled non-Hermitian  chains made out of basic circuit components e.g. capacitors, inductors, resistors, and INICs, and show that the NHSE bulk state accumulation can be induced or suppressed by varying the a.c. driving frequency of the circuit.  Interestingly, the coupled chain system exhibits NHSE state localization only when the driving frequency exceeds some critical frequency. It is also possible to tune the frequency so as to suppress the NHSE and restore the conventional BBC, even when the individual constituent non-Hermitian chains can  still host NHSE. In addition, at the high-frequency limit, the skin mode solutions of the coupled system become the superimposition of the skin solutions for the individual chains, which indicates an absence of hybridization between the two chains. In summary, our proposal which leverages on the tunability and accessibility of the TE platform, allows for the ready switching and modulation of the NHSE in a coupled system merely via frequency tuning, without any changes to the experimental circuit set-ups or component values.
  
\section{Results}
\begin{figure*}[htp!]
  \centering
    \includegraphics[width=0.9\textwidth]{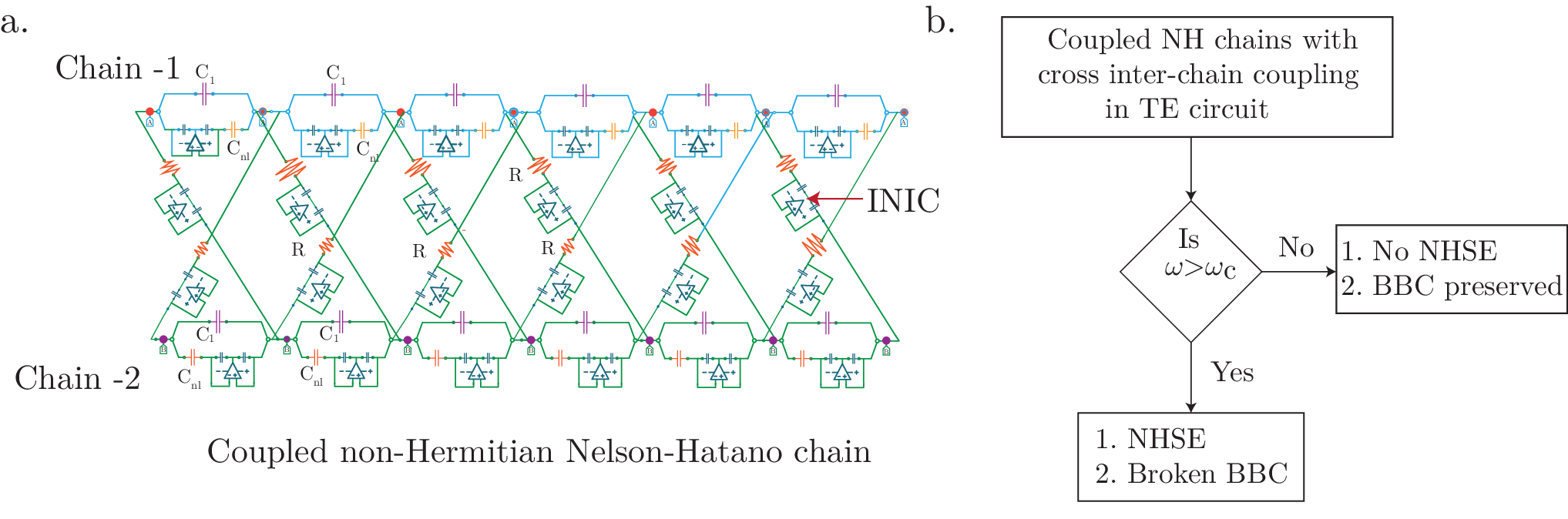}
  \caption{(a) A schematic diagram of the coupled non-Hermitian Hatano-Nelson TE chains. Each node is connected to its neighbors on the same chain by a capacitor $C_1$ with a coupling asymmetry of $\pm C_n$ depending on the coupling direction. Each node in a chain is also coupled to the neigbouring nodes on the other chain via a resistor passing through a INIC. The resistive coupling give rise to the frequency dependent terms in the Laplacian matrix. (b) Manifestation of frequency-controlled breakup and recovery of the bulk boundary correspondence (BBC) and non-Hermitian skin effect (NHSE).}
  \label{fig1}
\end{figure*}  
\begin{figure*}[htp!]
  \centering
    \includegraphics[width=0.8\textwidth]{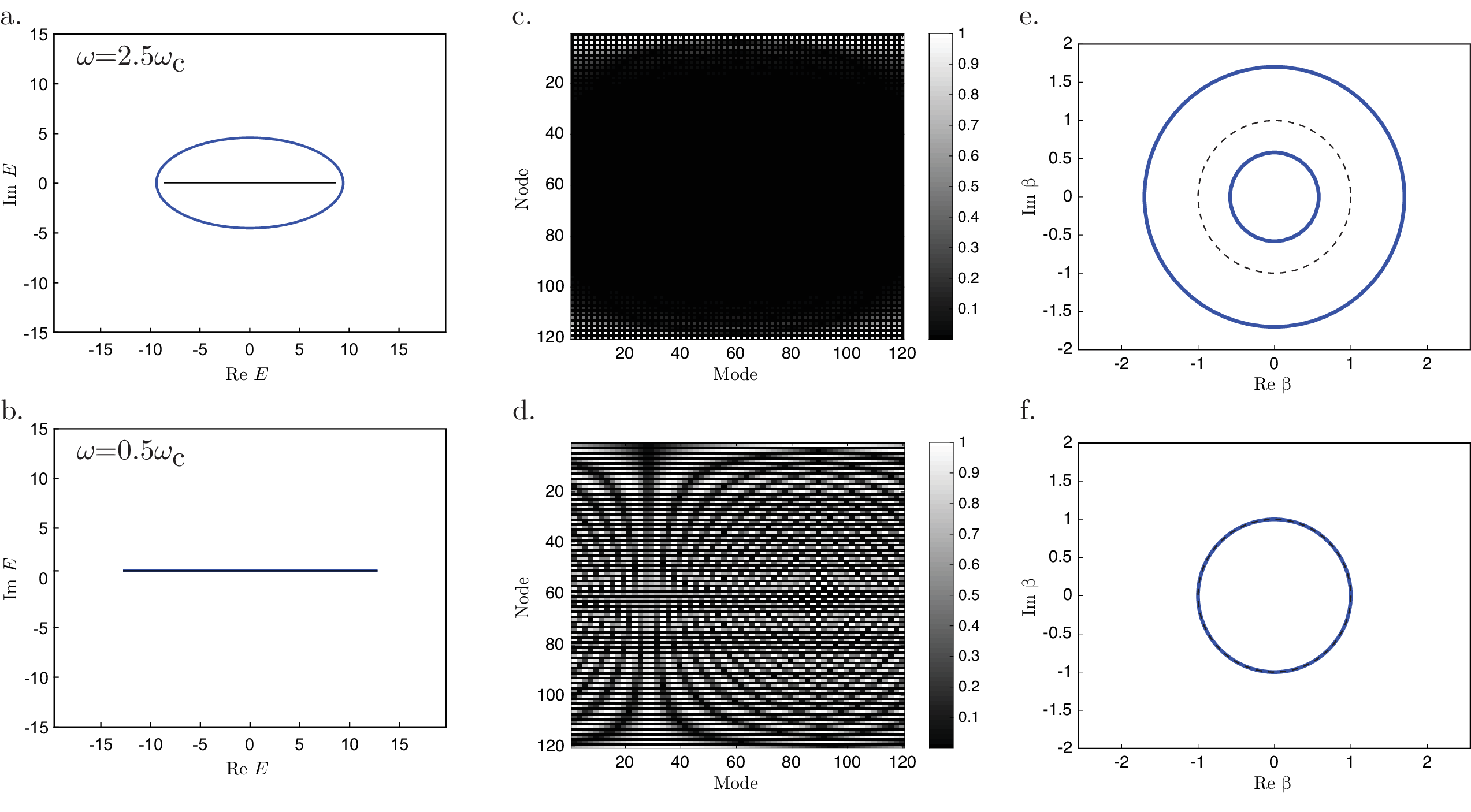}
  \caption{Frequency-controlled non-Hermitian characteristics in the coupled system shown in Fig. \ref{fig1}a. (a),(b) Calculated eigenvalue ($E$) spectra under OBC (thick blue lines) and PBC (thinner black lines) at a frequency (a) greater (i.e., $\omega=2.5 \omega_{\mathrm{c}}$) and (b) less (i.e., $\omega=0.5 \omega_{\mathrm{c}}$) than the critical frequency. (c),(d) The eigenmode voltage distributions at (c) $\omega=2.5 \omega_{\mathrm{c}}$ and (d) $\omega=0.5 \omega_{\mathrm{c}}$ for a composite system with 40 unit cells. The presence (absence) of extensive eigenstates localization at the boundaries at  $\omega > \omega_{\mathrm{c}}$ in (c) for $\omega > \omega_{\mathrm{c}}$ (in (d) for $\omega < \omega_{\mathrm{c}}$) signifies the appearance (disappearance) of NHSE. (e),(f) The distribution of the $\beta$ values present in the eigenmodes on the complex $\beta$ plane at driving frequencies (e) larger  and (f) smaller frequency than the critical frequency, respectively. When $\omega > \omega_{\mathrm{c}}$, the $\beta$ values fall on two circles corresponding to $\beta^{(1)}_\pm$ and $\beta^{(2)}_\pm$, which are bigger and smaller than the complex unit circle (dotted line), respectively. When $\omega < \omega_c$, the $\beta$ values falls on the unit circle (dotted line). Common parameters used: $C_1=4.7$ $\mu $F, $C_n=1.5$ $\mu$F and $R=150$ $\Omega$.}
  \label{fig2}
\end{figure*}  

%\begin{figure*}[htp!]
%  \centering
%    \includegraphics[width=0.5\textwidth]{Fig3.eps}
%  \caption{ Skin decay length of eigenstates in the coupled system as a function of frequency. The skin length becomes very large at $\omega< \omega_{\mathrm{c}}$, indicating the absence of NHSE. There is a sharp transition in the $\xi_{\mathrm{skin}}$ when system frequency matches with the critical frequency. The skin length exponentially decays when $\omega > \omega_{\mathrm{c}}$ and becomes saturated at very large frequency limit $\omega \gg \omega_{\mathrm{c}}$ (see inset). Common parameters used: $C_1=4.7$ $\mu $F, $C_n=1.5$ $\mu$F and $R=150$ $\Omega$.}
%  \label{fig3}
%\end{figure*}  
%\section{Methods}
\subsection{Derivation of frequency-induced NHSE}
In this section, we derive the conditions a.c. frequency drive for the emergence of the NHSE in the simple TE circuit shown in Fig. \ref{fig1}a. The circuit models a coupled system of two 1D non-Hermitian Hatano-Nelson chain with asymmetric next-neighbor couplings (i.e., $C_1 \pm C_n$). The two chains are connected to each other via frequency-dependent lossy and gain terms corresponding to positive and negative resistances realized using INICs.  A TE circuit under a mono-frequency drive at an angular frequency $\omega$ can be mathematically described by its Laplacian matrix $J$, which is obtained by applying Kirchoff's current law at each node \cite{lee2018topolectrical, rafi2020topoelectrical}. For convenience, we introduce $H \equiv -i J/\omega$. Under PBC, $H_{\mathrm{PBC}}$ can be written as 
\begin{widetext}
\begin{equation}
 H_{\mathrm{PBC}}^{\mathrm{coupled}}=\begin{pmatrix}
(C_1+C_n) e^{i k_x}+(C_1-C_n) e^{-i k_x} & -\frac{i}{\omega R}( e^{i k_x}- e^{-i k_x})\\ -\frac{i}{\omega R}( e^{i k_x}- e^{-i k_x})  & (C_1-C_n) e^{i k_x}+(C_1+C_n) e^{-i k_x}
\end{pmatrix},
\label{Ham1}
\end{equation}
\end{widetext}
where $C_n$ is the coupling asymmetry in the two non-Hermitian chains and $R$ the resistive coupling between the two chains. The sign of the resistive couplings is controlled using INICs where a resistive coupling passing through the positive (negative) terminal of the operational amplifier in an INIC segment represents a lossy (gainy) coupling, respectively (see Fig. \ref{fig1}a). The main findings of our study are outlined in the flowchart in Fig. \ref{fig1}b.  

The PBC eigenvalue spectrum (which we refer to subsequently as the admittance spectrum and denote the eigenvalues as $E$) of Eq. \eqref{Ham1} is given by 
\begin{equation}
E_{\mathrm{PBC}}=2 C_1 \cos k_x \pm \frac{2}{\omega R} \sqrt{1- (\omega R C_n)^2} \sin k_x.
\label{eq2}
\end{equation}
Eq. \eqref{eq2} implies that the admittance spectrum can be modified simply by tuning the AC driving frequency. Explicitly, Eq. \eqref{eq2} defines a critical frequency of $ \omega_{\mathrm{c}} = \frac{1}{R C_n} $ at which any driving frequency $\omega$ greater than $\omega_{\mathrm{c}}$ results in a complex PBC spectrum (see Fig. \ref{fig2}a). In contrast, the admittance spectrum assumes only real values when $\omega < \omega_{\mathrm{c}}$, as shown in Fig. \ref{fig2}b. 

We next turn to the behavior of the finite system and consider its OBC spectrum. The Hamiltonian for a system comprising of  non-Hermitian constituent subsystems can be written in the generic form of  
\begin{equation}
H_{\mathrm{coupled}} (\beta)= \begin{pmatrix}
\lambda_{\mathrm{chain 1}} (\beta) & \alpha (\omega,\beta) \\ 
\alpha (\omega,\beta) & \lambda_{\mathrm{chain 2}} (\beta)
\end{pmatrix}
\label{eq3}
\end{equation}
where $\lambda_{\mathrm{chain} i} (\beta)$ is ($-i/\omega$ times) the $\beta$-dependent Laplacian of the $i$th of two chains coupled via a frequency- and $\beta$-dependent term $\alpha (\omega,\beta)$. Here, $\beta$ is the non-Bloch factor, which takes the form of $\beta \to e^{k+i \kappa}$ with $\kappa$ denoting the amplification or attenuation factor of the eigenmode localization. The characteristic polynomial of $H_{\mathrm{coupled}} – E_n \mathbf{I}_2$, where $E_n$ is a generic eigenvalue and $\mathbf{I}_2$ is the 2 by 2 identity matrix, can be expressed as 
\begin{equation}
f_{\mathrm{ch}} (\omega, \beta )= (\lambda_{\mathrm{chain 1}} (\beta)-E_n)(\lambda_{\mathrm{chain 2}} (\beta)-E_n) -\alpha (\omega,\beta)^2.
\label{eq4}
\end{equation}

Because the inter-chain coupling gives rise to frequency- and $\beta$-dependent terms in the characteristic polynomial (see Eq. \eqref{eq4}), the skin mode solutions of the coupled system will be modified greatly from that of the decoupled system. For our TE model, the terms in Eq. \eqref{eq4} take the explicit form of
\begin{equation}
\begin{aligned} 
\lambda_{\mathrm{chain 1}} &=  C_1 \left(\beta+ \frac{1}{\beta}\right)+C_n \left(\beta - \frac{1}{\beta}\right), \\ 
\lambda_{\mathrm{chain 2}} &=  C_1 \left(\beta+ \frac{1}{\beta}\right)-C_n \left(\beta - \frac{1}{\beta}\right), \\ 
\alpha_(\omega,\beta) &= \frac{i}{\omega R} (\beta^{-1}-\beta).
\end{aligned}
\label{eq5}
\end{equation}
They collectively give the following characteristic equation 
\begin{equation}
a_4 \beta^4+a_3 \beta^3+a_2 \beta^2+a_1 \beta+ a_0=0
\label{eq6}
\end{equation}
with 
\begin{align*}
a_4 &= C_1^2-C_n^2 + (\omega R)^{-2} ,\\
a_3 &= - 2 E_n C_1, \\
a_2 &= E_n^2+2C_n^2+C_1^2 -2 (\omega R)^{-2}, \\
a_1 &= -2 E_n C_1, \\
a_0 &= C_1^2-C_n^2 - (\omega R)^{-2}. \\
\end{align*}
The characteristic equation in Eq. \eqref{eq6} can be factorized as the product of two quadratic equations $f_{\mathrm{ch}} (\omega, \beta )=f_{\mathrm{ch1}} (\omega, \beta ) f_{\mathrm{ch2}} (\omega, \beta )$ where
\begin{widetext}
\begin{align}
f_{\mathrm{ch1}} (\omega, \beta ) &= C_1 (1+ \beta^2)+ \sqrt{C_n^2-(\omega R)^{-2}}(-1+ \beta^2)-E_n \beta, \\
f_{\mathrm{ch2}} (\omega, \beta ) &= C_1 (1+ \beta^2)- \sqrt{C_n^2-(\omega R)^{-2}}(-1+ \beta^2)-E_n \beta.
\label{eq7}
\end{align}
\end{widetext}
Interestingly, the two independent quadratic equations in Eq. \eqref{eq7} satisfy $f_{\mathrm{ch1}} (\omega, \beta )=f_{\mathrm{ch2}} (\omega, \frac{1}{\beta} )$. This reflects the mutually reciprocal relationship between the skin mode solutions of the individual decoupled non-Hermitian chains in our system.

Equating $f_{\mathrm{ch1}}$ and $f_{\mathrm{ch2}}$ to zero to solve for the respective non-Bloch factors of chains 1 and 2, for a given $E_n$ and $\omega$, we obtain the four solutions 
\begin{align}
\beta^{(1)}_{\pm} &= \frac{E_n \pm \sqrt{E_n^2-4(C_1^2-C_n^2+ (\omega R)^{-2})}}{2(C_1-\sqrt{C_n^2- (\omega R)^{-2}})} \label{beta1}, \\ 
\beta^{(2)}_{\pm} &= \frac{E_n \pm \sqrt{E_n^2-4(C_1^2-C_n^2+ (\omega R)^{-2})}}{2(C_1+\sqrt{C_n^2- (\omega R)^{-2}})} \label{beta2}. 
\end{align}

Remarkably, by substituting the expressions for $\beta^{(j)}_\pm$ in Eqs. \eqref{beta1} and \eqref{beta2} into Eq. \eqref{Ham1}, it can be found that the eigenspinor $|v\rangle$ of $H(k=-I \log \beta^{(j)}_\pm)$ takes the form of 
\begin{equation}
	|v\rangle \propto \begin{pmatrix} 1 \\ -i R(C_n + \sqrt{C_n^2 – (\omega R)^{-2}}) \end{pmatrix}
\end{equation}
for all four $\beta^{(j)}_\pm$s, $j=(1,2)$. 

In particular, when $E_n$ is purely real and $\sqrt{E_n^2-4(C_1^2-C_n^2+ (\omega R)^{-2})}$ purely imaginary, the $\beta^{(j)}_\pm$s form a complex conjugate pair for a given $j=1,2$. In this case, the $\beta^{(j)}_\pm$s can be written as $\beta^{(j)}_\pm = \exp(\kappa^{(j)})\exp(\pm ik)$ where $\kappa^{(j)} = \mathrm{Ln} |\beta^{(j)}_\pm|$. (Note that $|\beta^{(j)}_+|= |\beta^{(j)}_-|$ when the $\beta^{(j)}_\pm$s form a complex conjugate pair), and $k=\mathrm{arg}(\beta^{(j)}_+)$ (since $\mathrm{arg}(\beta^{(1)}_+) = \mathrm{arg}(\beta^{(2)}_+)= -\mathrm{arg}(\beta^{(1)}_-)$ when the $\beta^{(j)}_\pm$s form a complex conjugate pair). An OBC eigenmode for a system that spans from $x=1$ to $x=N$ can then be written as 
\begin{equation}
	|\psi^{(j)}\rangle(x)= |v\rangle \exp(\kappa^{(j)}x) (\exp(ikx) – \exp(-ikx)) \label{psijx} 
\end{equation}
provided that $k(N+1)$ is an integer multiple of $\pi$ to satisfy the boundary condition that the eigenmode disappears at the $(N+1)$th unit cell beyond the right-most unit cell of the system.  

Note that this scenario differs from that of the conventional picture of the generalized Brillouin zone (GBZ) \cite{helbig2020generalized,kunst2018biorthogonal,yang2020non}  in which the OBC eigenstates in the thermodynamic limit occur at the eigenenergies where the middle $|\beta|$ values have the same moduli. In general, a two-by-two Hamiltonian with nearest-unit cell coupling requires the OBC eigenstate at a given eigenenergy $E_n$ to be constructed as a linear combination of all four periodic energy eigenstates to satisfy the four conditions corresponding to  the disappearance of the wavefunction  immediately beyond the left and right boundaries of the system when the eigenstates have different eigenspinors. In this case, the second and third largest $|\beta|$s of the system are required to have the same values so that their contributions can cancel out beyond both boundaries of the system in the thermodynamic limit. By contrast, for our present system, where all four periodic eigenmodes for a given $E_n$ have the same eigenspinor $|v\rangle$ ,  this allows $|v\rangle$ to be factored out so that an OBC eigenstate can be constructed from the linear combination of only two periodic eigenmodes, as per Eq. \eqref{psijx}, rather than all four of them.

The condition that $\sqrt{E_n^2-4(C_1^2-C_n^2+ (\omega R)^{-2})}$ is imaginary for the validity of Eq. \ref{psijx} then implies that $E_n^{\mathrm{OBC}}$, the eigenenergies that fulfill this condition, would satisfy
\begin{equation}
|E_n^{\mathrm{OBC}}|<2\sqrt{C_1^2-C_n^2+(\omega R)^{-2}}. 
\label{eq10}
\end{equation}

Note that the condition in Eq. \eqref{eq10} is valid only when $( \omega R)^{-2} > C_n^2 – C_1^2$. This can be ensured for all real values of $R$ and $\omega$ by choosing $C_1 > C_n$ for positive $C_1$ and $C_n$, as we do here. When this condition is fulfilled, the considerations leading to Eq. \eqref{psijx} imply that there are two eigenmodes at each value of $E_n^{\mathrm{OBC}}$ corresponding to $j=1$ and $j=2$ in Eq. \eqref{psijx} where, from Eqs. \eqref{beta1} and \eqref{beta2},
\begin{align}
	\kappa^{(1)} &= \left|\frac{C_1 - \sqrt{C_n^2 – (\omega R)^{-2}} }{C_1 + \sqrt{C_n^2 - (\omega R)^{-2}}}\right|, \label{kappa1} \\ 
	\kappa^{(2)} &= \left|\frac{C_1 + \sqrt{C_n^2 – (\omega R)^{-2}} }{C_1 - \sqrt{C_n^2 - (\omega R)^{-2}}}\right| \label{kappa2}.
\end{align} 

Notice that $\kappa^{(1)}$ and $\kappa^{(2)}$ are reciprocals of each other.  When $(\omega R)^{-2} > C_n^2$, or in other words when $\omega < \omega_{\mathrm{c}}$, $C_n^2 – (\omega R)^{-2} < 0$, in which case $\sqrt{C_n^2 – (\omega R)^{-2}}$ can be written as $i\lambda$ where $\lambda$ is real. Then, the numerator and denominator of $\kappa^{(j)}$ in Eqs. \ref{kappa1} and \ref{kappa2} become complex conjugates of each other, and $\kappa^{(j)}=1$. This corresponds to an absence of the NHSE. In contrast, when $\omega > \omega_{\mathrm{c}}$, $0 < \kappa^{(1)} \leq 1$ and $\kappa^{(2)} > 1$, we have a situation whereby the OBC eigenmodes exhibit a bipolar localization with the  $\kappa^{(1)}$ eigenmodes being localized nearer the left edge and the $\kappa^{(2)}$ eigenmodes nearer the right edge.

% nb there isn’t a CNHSE in this system because the GBZ of the composite system takes the same form as the GBZ of the component subsystems, i.e., a line along the real axis. We therefore need to remove references to the GBZ. Ah probably can replace the GBZ in c, d with the PBC spectrum. 

We illustrate these trends in an exemplary system in Fig. \ref{fig2}, which shows the effect of the driving frequency on the eigenvalue spectra, eigenmode spatial distribution, and  $\beta$ values under OBC of the composite system. When the driving frequency exceeds $\omega_{\mathrm{c}}$, the OBC and PBC eigenvalue spectra differ from each other with the former taking the form of an ellipse on the complex plane (Fig. \ref{fig2}a)  while the latter a straight line along the real axis (Fig. \ref{fig2}b). In the plot of the spatial distribution of the eigenmodes, there is  evident localization of the eigenmodes at both edges of the system (Fig. \ref{fig2}c). This bipolar localization is also in line with the plot of the eigenenergies over the complex $\beta$ plane (Fig. \ref{fig2}e), where the $\beta$ values fall on circles that differ from the dotted unit circle (one within and one encompassing it). In contrast, when the angular frequency is set at a value less than $\omega_{\mathrm{c}}$, the eigenvalue spectrum collapses to a line along the real axis under both PBC and OBC (see Fig. \ref{fig2}b). The bulk eigenstates are no longer localized at a particular edge of the system (Fig. \ref{fig2}d) even though each individual chain still retains its non-reciprocal coupling. This signifies the disappearance of the NHSE in the system, which can also be seen from the fact that $\beta$ values in the OBC eigenvalue spectrum now coincide with the unit circle (Fig. \ref{fig2}f). Thus, the dissimilar and identical OBC-PBC eigenvalue spectra at $\omega > \omega_{\mathrm{c}}$ and $\omega < \omega_{\mathrm{c}}$, respectively, correspond to the emergence and disappearance of the NHSE in our coupled system and the concomitant breakdown and restoration of the BBC in the coupled system.

\begin{figure}[htp!]
  \centering
    \includegraphics[width=0.48\textwidth]{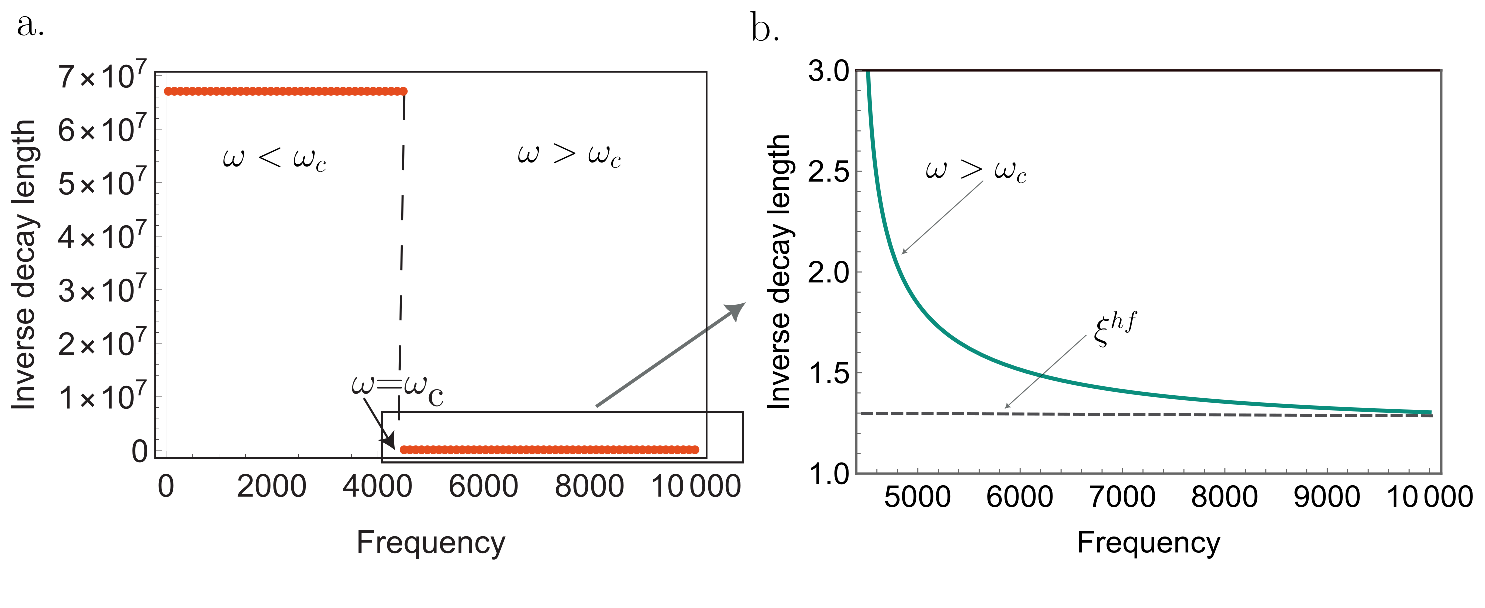}
  \caption{a. The skin decay length of bulk eigenstates  in the coupled system under study as a function of frequency $\omega$. The skin length becomes very large at $\omega< \omega_c$, indicating the absence of NHSE. There is a sharp transition in the $\xi_{skin}$ when system frequency matches the critical frequency $\omega_c$. b. The skin length exponentially decays when $\omega > \omega_c$ and becomes saturated at very large frequency limit $\omega \gg \omega_c$. Common parameters used: $C_1=4.7$ $\mu $F, $C_n=1.5$ $\mu$F and $R=150$ $\Omega$.}
  \label{fig3}
\end{figure}  

Likewise, the frequency of the ac drive can also be used to modulate the decay length within the coupled chain system. To obtain the frequency dependence of the decay length of the coupled chains, we first define the solutions to Eq. \ref{eq7} as $\beta_1$ and $\beta_2$. Along with the condition $|\beta_1|=|\beta_2|$, we then find the following expression for  decay length ($\xi_{skin}$) of the coupled chains to be
\begin{equation}
\xi_{skin}= 1/\sqrt{|\beta_1||\beta_2|}=2 \left(\log \left| \frac{C_1-\sqrt{C_n^2-\omega^{-2} R^{-2}}}{C_1+\sqrt{C_n^2-\omega^{-2} R^{-2}}} \right|\right)^{-1}
\label{eq11}
\end{equation}
The above expression reveals the frequency dependence of the decay length $\xi_{skin}$ of the all eigenmodes of our coupled system . The variation of $\xi_{skin}$ as a function of $\omega$ is plotted  in Fig. \ref{fig3}. For $\omega < \omega_c$, the decay length of the eigenstates approaches to a very large value (ideally infinity), suggesting that no eigenstates are localized in  that frequency range (see Fig. \ref{fig3}a). This can be explained by considering Eq. \ref{eq11}, where $\sqrt{C_n^2-\omega^{-2} R^{-2}}$ becomes purely imaginary at $\omega < \omega_c$ and can be expressed as $\sqrt{C_n^2-\omega^{-2} R^{-2}}= i \lambda$ where $\lambda$ is some real number. Hence the non-Bloch factor is reduced to $|\beta|=\sqrt{\left|\frac{C_1+i \lambda}{C_1-i \lambda}\right|}=1$, so that the decay length of the eigenmodes goes to $\xi_{skin}=\frac{1}{\log |\beta|}=\infty $. On the contrary,  for the case of $\omega > \omega_c$, the decay length is finite as the  term $\sqrt{C_n^2-\omega^{-2} R^{-2}}$ in Eq. \ref{eq11} is real, indicating the exponential decay of the eigenstates that are localized at the open boundaries (see Fig. \ref{fig3}b). Notice that there is a sharp transition of the $\xi_{skin}$, when the angular frequency matches the critical frequency of the TE circuit (see Fig. \ref{fig3}).

\subsection{Non-Hermitian characteristics at high-frequency limit}
In the high-frequency limit (i.e., $\omega \to \infty$), the off-diagonal terms in Eq. \eqref{Ham1} tend to 0. The characteristic equation in the high-frequency limit is then given by
\begin{align}
&(C_1 (1+ \beta^2)+ C_n(-1+ \beta^2)-E_n \beta)\times \nonumber \\ & (C_1 (1+ \beta^2)- C_n(-1+ \beta^2)-E_n \beta)=0
\label{eq13} 
\end{align}

The overall skin mode solution is hence given by the product of the skin mode solutions of the individual chains, and the decay lengths of  the bulk modes in the high-frquency limit, $\xi^{\mathrm{hf}} \equiv \log \kappa $, are given by 
\begin{equation}
\xi^{\mathrm{hf}}= \left( \log \left| \frac{C_1-C_n}{C_1+C_n}\right| \right)^{ \pm 1}.
\label{eq14}
\end{equation}

Eq. \eqref{eq14} shows that eigenstate distribution becomes independent of frequency and is solely dependent on the coupling asymmetry in the individual chain at the high-frequency limit. In other words, the decay length reaches an asymptotic value at high frequencies as shown in Fig. \ref{fig3}b.

Thus, the frequency-controlled breaking of the BBC in the coupled system occurs as long as the non-Hermiticity in the component chains are maintained through coupling asymmetry in the respective chains. If the coupling asymmetry is turned off, i.e., $C_n=0$, the non-Bloch factor becomes unity (e.g., $|\beta|=1$) independent of the driving frequency. As a result, the decay length of the eigenmode distribution becomes infinity and there is no skin mode localization. 

\subsection{Conclusion}
In summary, we proposed a non-Hermitian coupled TE system comprising two non-reciprocal capacitively-coupled Hatano-Nelson chains with mutually inverse decay lengths connected via resistive coupling. The resistive coupling results in an ac  frequency dependence of the characteristic equation. We analyze the effect of the driving frequency on the NHSE both analytically and numerically. The eigenmodes exhibit the NHSE and associated non-Hermitian characteristics only when the frequency  exceeds a critical value $\omega_{\mathrm{c}}$ . Analytically, this is due to  the hybridization of the skin mode solutions for the two non-Hermitian chains via the frequency-dependent term in the characteristic equation. In contrast, when the driving frequency  $\omega < \omega_{\mathrm{c}}$, the NHSE is suppressed and the conventional BBC restored even though the individual uncoupled chains still exhibit NHSE. In addition, in the high-frequency limit (i.e., $\omega >> \omega_{\mathrm{c}}$), the overall skin decay length of the coupled system becomes independent of the frequency and system size as the skin mode solutions are given simply by the combination of the NHSE solutions for the individual chains. The proposed method of frequency tuning provides a simple and effective means to modulate and investigate non-Hermitian phenomena without any change to the topolectrical circuit set-up.

\subsection*{acknowledgment}
This work is supported by the Ministry of Education (MOE) Tier-II Grant No. MOE-T2EP50121-0014 (NUS Grant No. A-8000086-01-00) and MOE Tier-I FRC Grant (NUS Grant No. A-8000195-01-00).
  
%\bibliographystyle{apsrev4-2}
%\bibliography{freq_dep_CNHSE_v1}

\begin{thebibliography}{59}%
\makeatletter
\providecommand \@ifxundefined [1]{%
 \@ifx{#1\undefined}
}%
\providecommand \@ifnum [1]{%
 \ifnum #1\expandafter \@firstoftwo
 \else \expandafter \@secondoftwo
 \fi
}%
\providecommand \@ifx [1]{%
 \ifx #1\expandafter \@firstoftwo
 \else \expandafter \@secondoftwo
 \fi
}%
\providecommand \natexlab [1]{#1}%
\providecommand \enquote  [1]{``#1''}%
\providecommand \bibnamefont  [1]{#1}%
\providecommand \bibfnamefont [1]{#1}%
\providecommand \citenamefont [1]{#1}%
\providecommand \href@noop [0]{\@secondoftwo}%
\providecommand \href [0]{\begingroup \@sanitize@url \@href}%
\providecommand \@href[1]{\@@startlink{#1}\@@href}%
\providecommand \@@href[1]{\endgroup#1\@@endlink}%
\providecommand \@sanitize@url [0]{\catcode `\\12\catcode `\$12\catcode
  `\&12\catcode `\#12\catcode `\^12\catcode `\_12\catcode `\%12\relax}%
\providecommand \@@startlink[1]{}%
\providecommand \@@endlink[0]{}%
\providecommand \url  [0]{\begingroup\@sanitize@url \@url }%
\providecommand \@url [1]{\endgroup\@href {#1}{\urlprefix }}%
\providecommand \urlprefix  [0]{URL }%
\providecommand \Eprint [0]{\href }%
\providecommand \doibase [0]{https://doi.org/}%
\providecommand \selectlanguage [0]{\@gobble}%
\providecommand \bibinfo  [0]{\@secondoftwo}%
\providecommand \bibfield  [0]{\@secondoftwo}%
\providecommand \translation [1]{[#1]}%
\providecommand \BibitemOpen [0]{}%
\providecommand \bibitemStop [0]{}%
\providecommand \bibitemNoStop [0]{.\EOS\space}%
\providecommand \EOS [0]{\spacefactor3000\relax}%
\providecommand \BibitemShut  [1]{\csname bibitem#1\endcsname}%
\let\auto@bib@innerbib\@empty
%</preamble>
\bibitem [{\citenamefont {Rafi-Ul-Islam}\ \emph
  {et~al.}(2021{\natexlab{a}})\citenamefont {Rafi-Ul-Islam}, \citenamefont
  {Siu},\ and\ \citenamefont {Jalil}}]{rafi2021non}%
  \BibitemOpen
  \bibfield  {author} {\bibinfo {author} {\bibfnamefont {S.}~\bibnamefont
  {Rafi-Ul-Islam}}, \bibinfo {author} {\bibfnamefont {Z.~B.}\ \bibnamefont
  {Siu}},\ and\ \bibinfo {author} {\bibfnamefont {M.~B.}\ \bibnamefont
  {Jalil}},\ }\href@noop {} {\bibfield  {journal} {\bibinfo  {journal} {New J.
  Phys.}\ }\textbf {\bibinfo {volume} {23}},\ \bibinfo {pages} {033014}
  (\bibinfo {year} {2021}{\natexlab{a}})}\BibitemShut {NoStop}%
\bibitem [{\citenamefont {Rafi-Ul-Islam}\ \emph
  {et~al.}(2022{\natexlab{a}})\citenamefont {Rafi-Ul-Islam}, \citenamefont
  {Siu}, \citenamefont {Sahin}, \citenamefont {Lee},\ and\ \citenamefont
  {Jalil}}]{PhysRevResearch.4.043108}%
  \BibitemOpen
  \bibfield  {author} {\bibinfo {author} {\bibfnamefont {S.~M.}\ \bibnamefont
  {Rafi-Ul-Islam}}, \bibinfo {author} {\bibfnamefont {Z.~B.}\ \bibnamefont
  {Siu}}, \bibinfo {author} {\bibfnamefont {H.}~\bibnamefont {Sahin}}, \bibinfo
  {author} {\bibfnamefont {C.~H.}\ \bibnamefont {Lee}},\ and\ \bibinfo {author}
  {\bibfnamefont {M.~B.~A.}\ \bibnamefont {Jalil}},\ }\href
  {https://doi.org/10.1103/PhysRevResearch.4.043108} {\bibfield  {journal}
  {\bibinfo  {journal} {Phys. Rev. Res.}\ }\textbf {\bibinfo {volume} {4}},\
  \bibinfo {pages} {043108} (\bibinfo {year} {2022}{\natexlab{a}})}\BibitemShut
  {NoStop}%
\bibitem [{\citenamefont {Gao}\ \emph {et~al.}(2020)\citenamefont {Gao},
  \citenamefont {Willatzen},\ and\ \citenamefont
  {Christensen}}]{gao2020anomalous}%
  \BibitemOpen
  \bibfield  {author} {\bibinfo {author} {\bibfnamefont {P.}~\bibnamefont
  {Gao}}, \bibinfo {author} {\bibfnamefont {M.}~\bibnamefont {Willatzen}},\
  and\ \bibinfo {author} {\bibfnamefont {J.}~\bibnamefont {Christensen}},\
  }\href@noop {} {\bibfield  {journal} {\bibinfo  {journal} {Phys. Rev. Lett.}\
  }\textbf {\bibinfo {volume} {125}},\ \bibinfo {pages} {206402} (\bibinfo
  {year} {2020})}\BibitemShut {NoStop}%
\bibitem [{\citenamefont {Rafi-Ul-Islam}\ \emph
  {et~al.}(2022{\natexlab{b}})\citenamefont {Rafi-Ul-Islam}, \citenamefont
  {Sahin}, \citenamefont {Siu},\ and\ \citenamefont
  {Jalil}}]{rafi2022interfacial}%
  \BibitemOpen
  \bibfield  {author} {\bibinfo {author} {\bibfnamefont {S.}~\bibnamefont
  {Rafi-Ul-Islam}}, \bibinfo {author} {\bibfnamefont {H.}~\bibnamefont
  {Sahin}}, \bibinfo {author} {\bibfnamefont {Z.~B.}\ \bibnamefont {Siu}},\
  and\ \bibinfo {author} {\bibfnamefont {M.~B.}\ \bibnamefont {Jalil}},\
  }\href@noop {} {\bibfield  {journal} {\bibinfo  {journal} {Phys. Rev. Res.}\
  }\textbf {\bibinfo {volume} {4}},\ \bibinfo {pages} {043021} (\bibinfo {year}
  {2022}{\natexlab{b}})}\BibitemShut {NoStop}%
\bibitem [{\citenamefont {Lu}\ \emph {et~al.}(2021)\citenamefont {Lu},
  \citenamefont {Zhang},\ and\ \citenamefont {Franz}}]{lu2021magnetic}%
  \BibitemOpen
  \bibfield  {author} {\bibinfo {author} {\bibfnamefont {M.}~\bibnamefont
  {Lu}}, \bibinfo {author} {\bibfnamefont {X.-X.}\ \bibnamefont {Zhang}},\ and\
  \bibinfo {author} {\bibfnamefont {M.}~\bibnamefont {Franz}},\ }\href@noop {}
  {\bibfield  {journal} {\bibinfo  {journal} {Phys. Rev. Lett.}\ }\textbf
  {\bibinfo {volume} {127}},\ \bibinfo {pages} {256402} (\bibinfo {year}
  {2021})}\BibitemShut {NoStop}%
\bibitem [{\citenamefont {Rafi-Ul-Islam}\ \emph
  {et~al.}(2024{\natexlab{a}})\citenamefont {Rafi-Ul-Islam}, \citenamefont
  {Siu}, \citenamefont {Razo}, \citenamefont {Sahin},\ and\ \citenamefont
  {Jalil}}]{PhysRevB.110.045444}%
  \BibitemOpen
  \bibfield  {author} {\bibinfo {author} {\bibfnamefont {S.~M.}\ \bibnamefont
  {Rafi-Ul-Islam}}, \bibinfo {author} {\bibfnamefont {Z.~B.}\ \bibnamefont
  {Siu}}, \bibinfo {author} {\bibfnamefont {M.~S.~H.}\ \bibnamefont {Razo}},
  \bibinfo {author} {\bibfnamefont {H.}~\bibnamefont {Sahin}},\ and\ \bibinfo
  {author} {\bibfnamefont {M.~B.~A.}\ \bibnamefont {Jalil}},\ }\href
  {https://doi.org/10.1103/PhysRevB.110.045444} {\bibfield  {journal} {\bibinfo
   {journal} {Phys. Rev. B}\ }\textbf {\bibinfo {volume} {110}},\ \bibinfo
  {pages} {045444} (\bibinfo {year} {2024}{\natexlab{a}})}\BibitemShut
  {NoStop}%
\bibitem [{\citenamefont {Xu}\ and\ \citenamefont
  {Guo}(2022)}]{xu2022unconventional}%
  \BibitemOpen
  \bibfield  {author} {\bibinfo {author} {\bibfnamefont {J.}~\bibnamefont
  {Xu}}\ and\ \bibinfo {author} {\bibfnamefont {Y.}~\bibnamefont {Guo}},\
  }\href@noop {} {\bibfield  {journal} {\bibinfo  {journal} {New J. Phys.}\
  }\textbf {\bibinfo {volume} {24}},\ \bibinfo {pages} {053028} (\bibinfo
  {year} {2022})}\BibitemShut {NoStop}%
\bibitem [{\citenamefont {Rafi-Ul-Islam}\ \emph
  {et~al.}(2024{\natexlab{b}})\citenamefont {Rafi-Ul-Islam}, \citenamefont
  {Siu}, \citenamefont {Sahin}, \citenamefont {Razo},\ and\ \citenamefont
  {Jalil}}]{rafi2024twisted}%
  \BibitemOpen
  \bibfield  {author} {\bibinfo {author} {\bibfnamefont {S.}~\bibnamefont
  {Rafi-Ul-Islam}}, \bibinfo {author} {\bibfnamefont {Z.~B.}\ \bibnamefont
  {Siu}}, \bibinfo {author} {\bibfnamefont {H.}~\bibnamefont {Sahin}}, \bibinfo
  {author} {\bibfnamefont {M.~S.~H.}\ \bibnamefont {Razo}},\ and\ \bibinfo
  {author} {\bibfnamefont {M.~B.}\ \bibnamefont {Jalil}},\ }\href@noop {}
  {\bibfield  {journal} {\bibinfo  {journal} {Phys. Rev. B}\ }\textbf {\bibinfo
  {volume} {109}},\ \bibinfo {pages} {045410} (\bibinfo {year}
  {2024}{\natexlab{b}})}\BibitemShut {NoStop}%
\bibitem [{\citenamefont {Fan}\ \emph {et~al.}(2022)\citenamefont {Fan},
  \citenamefont {Liang}, \citenamefont {Li}, \citenamefont {Tsai},\ and\
  \citenamefont {Zhang}}]{fan2022emerging}%
  \BibitemOpen
  \bibfield  {author} {\bibinfo {author} {\bibfnamefont {Y.}~\bibnamefont
  {Fan}}, \bibinfo {author} {\bibfnamefont {H.}~\bibnamefont {Liang}}, \bibinfo
  {author} {\bibfnamefont {J.}~\bibnamefont {Li}}, \bibinfo {author}
  {\bibfnamefont {D.~P.}\ \bibnamefont {Tsai}},\ and\ \bibinfo {author}
  {\bibfnamefont {S.}~\bibnamefont {Zhang}},\ }\href@noop {} {\bibfield
  {journal} {\bibinfo  {journal} {ACS Photonics}\ } (\bibinfo {year}
  {2022})}\BibitemShut {NoStop}%
\bibitem [{\citenamefont {Gong}\ \emph {et~al.}(2018)\citenamefont {Gong},
  \citenamefont {Ashida}, \citenamefont {Kawabata}, \citenamefont {Takasan},
  \citenamefont {Higashikawa},\ and\ \citenamefont
  {Ueda}}]{gong2018topological}%
  \BibitemOpen
  \bibfield  {author} {\bibinfo {author} {\bibfnamefont {Z.}~\bibnamefont
  {Gong}}, \bibinfo {author} {\bibfnamefont {Y.}~\bibnamefont {Ashida}},
  \bibinfo {author} {\bibfnamefont {K.}~\bibnamefont {Kawabata}}, \bibinfo
  {author} {\bibfnamefont {K.}~\bibnamefont {Takasan}}, \bibinfo {author}
  {\bibfnamefont {S.}~\bibnamefont {Higashikawa}},\ and\ \bibinfo {author}
  {\bibfnamefont {M.}~\bibnamefont {Ueda}},\ }\href@noop {} {\bibfield
  {journal} {\bibinfo  {journal} {Phys. Rev. X}\ }\textbf {\bibinfo {volume}
  {8}},\ \bibinfo {pages} {031079} (\bibinfo {year} {2018})}\BibitemShut
  {NoStop}%
\bibitem [{\citenamefont {Bergholtz}\ \emph {et~al.}(2021)\citenamefont
  {Bergholtz}, \citenamefont {Budich},\ and\ \citenamefont
  {Kunst}}]{bergholtz2021exceptional}%
  \BibitemOpen
  \bibfield  {author} {\bibinfo {author} {\bibfnamefont {E.~J.}\ \bibnamefont
  {Bergholtz}}, \bibinfo {author} {\bibfnamefont {J.~C.}\ \bibnamefont
  {Budich}},\ and\ \bibinfo {author} {\bibfnamefont {F.~K.}\ \bibnamefont
  {Kunst}},\ }\href@noop {} {\bibfield  {journal} {\bibinfo  {journal} {Rev.
  Mod. Phys.}\ }\textbf {\bibinfo {volume} {93}},\ \bibinfo {pages} {015005}
  (\bibinfo {year} {2021})}\BibitemShut {NoStop}%
\bibitem [{\citenamefont {Leykam}\ \emph {et~al.}(2017)\citenamefont {Leykam},
  \citenamefont {Bliokh}, \citenamefont {Huang}, \citenamefont {Chong},\ and\
  \citenamefont {Nori}}]{leykam2017edge}%
  \BibitemOpen
  \bibfield  {author} {\bibinfo {author} {\bibfnamefont {D.}~\bibnamefont
  {Leykam}}, \bibinfo {author} {\bibfnamefont {K.~Y.}\ \bibnamefont {Bliokh}},
  \bibinfo {author} {\bibfnamefont {C.}~\bibnamefont {Huang}}, \bibinfo
  {author} {\bibfnamefont {Y.~D.}\ \bibnamefont {Chong}},\ and\ \bibinfo
  {author} {\bibfnamefont {F.}~\bibnamefont {Nori}},\ }\href@noop {} {\bibfield
   {journal} {\bibinfo  {journal} {Phys. Rev. Lett.}\ }\textbf {\bibinfo
  {volume} {118}},\ \bibinfo {pages} {040401} (\bibinfo {year}
  {2017})}\BibitemShut {NoStop}%
\bibitem [{\citenamefont {Rafi-Ul-Islam}\ \emph
  {et~al.}(2021{\natexlab{b}})\citenamefont {Rafi-Ul-Islam}, \citenamefont
  {Siu},\ and\ \citenamefont {Jalil}}]{rafi2021topological}%
  \BibitemOpen
  \bibfield  {author} {\bibinfo {author} {\bibfnamefont {S.}~\bibnamefont
  {Rafi-Ul-Islam}}, \bibinfo {author} {\bibfnamefont {Z.~B.}\ \bibnamefont
  {Siu}},\ and\ \bibinfo {author} {\bibfnamefont {M.~B.}\ \bibnamefont
  {Jalil}},\ }\href@noop {} {\bibfield  {journal} {\bibinfo  {journal} {Phys.
  Rev. B}\ }\textbf {\bibinfo {volume} {103}},\ \bibinfo {pages} {035420}
  (\bibinfo {year} {2021}{\natexlab{b}})}\BibitemShut {NoStop}%
\bibitem [{\citenamefont {Siu}\ \emph {et~al.}(2023)\citenamefont {Siu},
  \citenamefont {Rafi-Ul-Islam},\ and\ \citenamefont
  {Jalil}}]{siu2023terminal}%
  \BibitemOpen
  \bibfield  {author} {\bibinfo {author} {\bibfnamefont {Z.~B.}\ \bibnamefont
  {Siu}}, \bibinfo {author} {\bibfnamefont {S.}~\bibnamefont {Rafi-Ul-Islam}},\
  and\ \bibinfo {author} {\bibfnamefont {M.~B.}\ \bibnamefont {Jalil}},\
  }\href@noop {} {\bibfield  {journal} {\bibinfo  {journal} {Sci. Rep.}\
  }\textbf {\bibinfo {volume} {13}},\ \bibinfo {pages} {22770} (\bibinfo {year}
  {2023})}\BibitemShut {NoStop}%
\bibitem [{\citenamefont {Rafi-Ul-Islam}\ \emph
  {et~al.}(2024{\natexlab{c}})\citenamefont {Rafi-Ul-Islam}, \citenamefont
  {Siu}, \citenamefont {Razo},\ and\ \citenamefont
  {Jalil}}]{rafi2024saturation}%
  \BibitemOpen
  \bibfield  {author} {\bibinfo {author} {\bibfnamefont {S.}~\bibnamefont
  {Rafi-Ul-Islam}}, \bibinfo {author} {\bibfnamefont {Z.~B.}\ \bibnamefont
  {Siu}}, \bibinfo {author} {\bibfnamefont {M.~S.~H.}\ \bibnamefont {Razo}},\
  and\ \bibinfo {author} {\bibfnamefont {M.}~\bibnamefont {Jalil}},\
  }\href@noop {} {\bibfield  {journal} {\bibinfo  {journal} {ArXiv preprint
  arXiv:2406.19629}\ } (\bibinfo {year} {2024}{\natexlab{c}})}\BibitemShut
  {NoStop}%
\bibitem [{\citenamefont {Longhi}(2019)}]{longhi2019probing}%
  \BibitemOpen
  \bibfield  {author} {\bibinfo {author} {\bibfnamefont {S.}~\bibnamefont
  {Longhi}},\ }\href@noop {} {\bibfield  {journal} {\bibinfo  {journal} {Phys.
  Rev. Res.}\ }\textbf {\bibinfo {volume} {1}},\ \bibinfo {pages} {023013}
  (\bibinfo {year} {2019})}\BibitemShut {NoStop}%
\bibitem [{\citenamefont {Rafi-Ul-Islam}\ \emph
  {et~al.}(2022{\natexlab{c}})\citenamefont {Rafi-Ul-Islam}, \citenamefont
  {Siu}, \citenamefont {Sahin},\ and\ \citenamefont {Jalil}}]{rafi2022type}%
  \BibitemOpen
  \bibfield  {author} {\bibinfo {author} {\bibfnamefont {S.}~\bibnamefont
  {Rafi-Ul-Islam}}, \bibinfo {author} {\bibfnamefont {Z.~B.}\ \bibnamefont
  {Siu}}, \bibinfo {author} {\bibfnamefont {H.}~\bibnamefont {Sahin}},\ and\
  \bibinfo {author} {\bibfnamefont {M.~B.}\ \bibnamefont {Jalil}},\ }\href@noop
  {} {\bibfield  {journal} {\bibinfo  {journal} {Phys. Rev. B}\ }\textbf
  {\bibinfo {volume} {106}},\ \bibinfo {pages} {245128} (\bibinfo {year}
  {2022}{\natexlab{c}})}\BibitemShut {NoStop}%
\bibitem [{\citenamefont {Song}\ \emph {et~al.}(2019)\citenamefont {Song},
  \citenamefont {Yao},\ and\ \citenamefont {Wang}}]{song2019non}%
  \BibitemOpen
  \bibfield  {author} {\bibinfo {author} {\bibfnamefont {F.}~\bibnamefont
  {Song}}, \bibinfo {author} {\bibfnamefont {S.}~\bibnamefont {Yao}},\ and\
  \bibinfo {author} {\bibfnamefont {Z.}~\bibnamefont {Wang}},\ }\href@noop {}
  {\bibfield  {journal} {\bibinfo  {journal} {Phys. Rev. Lett.}\ }\textbf
  {\bibinfo {volume} {123}},\ \bibinfo {pages} {170401} (\bibinfo {year}
  {2019})}\BibitemShut {NoStop}%
\bibitem [{\citenamefont {Rafi-Ul-Islam}\ \emph
  {et~al.}(2022{\natexlab{d}})\citenamefont {Rafi-Ul-Islam}, \citenamefont
  {Siu}, \citenamefont {Sahin}, \citenamefont {Lee},\ and\ \citenamefont
  {Jalil}}]{rafi2022unconventional}%
  \BibitemOpen
  \bibfield  {author} {\bibinfo {author} {\bibfnamefont {S.}~\bibnamefont
  {Rafi-Ul-Islam}}, \bibinfo {author} {\bibfnamefont {Z.~B.}\ \bibnamefont
  {Siu}}, \bibinfo {author} {\bibfnamefont {H.}~\bibnamefont {Sahin}}, \bibinfo
  {author} {\bibfnamefont {C.~H.}\ \bibnamefont {Lee}},\ and\ \bibinfo {author}
  {\bibfnamefont {M.~B.}\ \bibnamefont {Jalil}},\ }\href@noop {} {\bibfield
  {journal} {\bibinfo  {journal} {Phys. Rev. Res.}\ }\textbf {\bibinfo {volume}
  {4}},\ \bibinfo {pages} {043108} (\bibinfo {year}
  {2022}{\natexlab{d}})}\BibitemShut {NoStop}%
\bibitem [{\citenamefont {Longhi}(2020)}]{longhi2020unraveling}%
  \BibitemOpen
  \bibfield  {author} {\bibinfo {author} {\bibfnamefont {S.}~\bibnamefont
  {Longhi}},\ }\href@noop {} {\bibfield  {journal} {\bibinfo  {journal} {Phys.
  Rev. B}\ }\textbf {\bibinfo {volume} {102}},\ \bibinfo {pages} {201103}
  (\bibinfo {year} {2020})}\BibitemShut {NoStop}%
\bibitem [{\citenamefont {Lee}\ and\ \citenamefont
  {Thomale}(2019)}]{lee2019anatomy}%
  \BibitemOpen
  \bibfield  {author} {\bibinfo {author} {\bibfnamefont {C.~H.}\ \bibnamefont
  {Lee}}\ and\ \bibinfo {author} {\bibfnamefont {R.}~\bibnamefont {Thomale}},\
  }\href@noop {} {\bibfield  {journal} {\bibinfo  {journal} {Phys. Rev. B}\
  }\textbf {\bibinfo {volume} {99}},\ \bibinfo {pages} {201103} (\bibinfo
  {year} {2019})}\BibitemShut {NoStop}%
\bibitem [{\citenamefont {Guo}\ \emph {et~al.}(2021)\citenamefont {Guo},
  \citenamefont {Liu}, \citenamefont {Zhao}, \citenamefont {Liu},\ and\
  \citenamefont {Chen}}]{guo2021exact}%
  \BibitemOpen
  \bibfield  {author} {\bibinfo {author} {\bibfnamefont {C.-X.}\ \bibnamefont
  {Guo}}, \bibinfo {author} {\bibfnamefont {C.-H.}\ \bibnamefont {Liu}},
  \bibinfo {author} {\bibfnamefont {X.-M.}\ \bibnamefont {Zhao}}, \bibinfo
  {author} {\bibfnamefont {Y.}~\bibnamefont {Liu}},\ and\ \bibinfo {author}
  {\bibfnamefont {S.}~\bibnamefont {Chen}},\ }\href@noop {} {\bibfield
  {journal} {\bibinfo  {journal} {Phys. Rev. Lett.}\ }\textbf {\bibinfo
  {volume} {127}},\ \bibinfo {pages} {116801} (\bibinfo {year}
  {2021})}\BibitemShut {NoStop}%
\bibitem [{\citenamefont {Rafi-Ul-Islam}\ \emph
  {et~al.}(2022{\natexlab{e}})\citenamefont {Rafi-Ul-Islam}, \citenamefont
  {Siu}, \citenamefont {Sahin}, \citenamefont {Lee},\ and\ \citenamefont
  {Jalil}}]{rafi2022system}%
  \BibitemOpen
  \bibfield  {author} {\bibinfo {author} {\bibfnamefont {S.}~\bibnamefont
  {Rafi-Ul-Islam}}, \bibinfo {author} {\bibfnamefont {Z.~B.}\ \bibnamefont
  {Siu}}, \bibinfo {author} {\bibfnamefont {H.}~\bibnamefont {Sahin}}, \bibinfo
  {author} {\bibfnamefont {C.~H.}\ \bibnamefont {Lee}},\ and\ \bibinfo {author}
  {\bibfnamefont {M.~B.}\ \bibnamefont {Jalil}},\ }\href@noop {} {\bibfield
  {journal} {\bibinfo  {journal} {Phys. Rev. B}\ }\textbf {\bibinfo {volume}
  {106}},\ \bibinfo {pages} {075158} (\bibinfo {year}
  {2022}{\natexlab{e}})}\BibitemShut {NoStop}%
\bibitem [{\citenamefont {Edvardsson}\ and\ \citenamefont
  {Ardonne}(2022)}]{edvardsson2022sensitivity}%
  \BibitemOpen
  \bibfield  {author} {\bibinfo {author} {\bibfnamefont {E.}~\bibnamefont
  {Edvardsson}}\ and\ \bibinfo {author} {\bibfnamefont {E.}~\bibnamefont
  {Ardonne}},\ }\href@noop {} {\bibfield  {journal} {\bibinfo  {journal} {Phys.
  Rev. B}\ }\textbf {\bibinfo {volume} {106}},\ \bibinfo {pages} {115107}
  (\bibinfo {year} {2022})}\BibitemShut {NoStop}%
\bibitem [{\citenamefont {Xiong}(2018)}]{xiong2018does}%
  \BibitemOpen
  \bibfield  {author} {\bibinfo {author} {\bibfnamefont {Y.}~\bibnamefont
  {Xiong}},\ }\href@noop {} {\bibfield  {journal} {\bibinfo  {journal} {J.
  Phys. Commun.}\ }\textbf {\bibinfo {volume} {2}},\ \bibinfo {pages} {035043}
  (\bibinfo {year} {2018})}\BibitemShut {NoStop}%
\bibitem [{\citenamefont {Jin}\ and\ \citenamefont {Song}(2019)}]{jin2019bulk}%
  \BibitemOpen
  \bibfield  {author} {\bibinfo {author} {\bibfnamefont {L.}~\bibnamefont
  {Jin}}\ and\ \bibinfo {author} {\bibfnamefont {Z.}~\bibnamefont {Song}},\
  }\href@noop {} {\bibfield  {journal} {\bibinfo  {journal} {Phys. Rev. B}\
  }\textbf {\bibinfo {volume} {99}},\ \bibinfo {pages} {081103} (\bibinfo
  {year} {2019})}\BibitemShut {NoStop}%
\bibitem [{\citenamefont {Xiao}\ \emph {et~al.}(2020)\citenamefont {Xiao},
  \citenamefont {Deng}, \citenamefont {Wang}, \citenamefont {Zhu},
  \citenamefont {Wang}, \citenamefont {Yi},\ and\ \citenamefont
  {Xue}}]{xiao2020non}%
  \BibitemOpen
  \bibfield  {author} {\bibinfo {author} {\bibfnamefont {L.}~\bibnamefont
  {Xiao}}, \bibinfo {author} {\bibfnamefont {T.}~\bibnamefont {Deng}}, \bibinfo
  {author} {\bibfnamefont {K.}~\bibnamefont {Wang}}, \bibinfo {author}
  {\bibfnamefont {G.}~\bibnamefont {Zhu}}, \bibinfo {author} {\bibfnamefont
  {Z.}~\bibnamefont {Wang}}, \bibinfo {author} {\bibfnamefont {W.}~\bibnamefont
  {Yi}},\ and\ \bibinfo {author} {\bibfnamefont {P.}~\bibnamefont {Xue}},\
  }\href@noop {} {\bibfield  {journal} {\bibinfo  {journal} {Nat. Phys.}\
  }\textbf {\bibinfo {volume} {16}},\ \bibinfo {pages} {761} (\bibinfo {year}
  {2020})}\BibitemShut {NoStop}%
\bibitem [{\citenamefont {Schindler}\ and\ \citenamefont
  {Prem}(2021)}]{schindler2021dislocation}%
  \BibitemOpen
  \bibfield  {author} {\bibinfo {author} {\bibfnamefont {F.}~\bibnamefont
  {Schindler}}\ and\ \bibinfo {author} {\bibfnamefont {A.}~\bibnamefont
  {Prem}},\ }\href@noop {} {\bibfield  {journal} {\bibinfo  {journal} {Phys.
  Rev. B}\ }\textbf {\bibinfo {volume} {104}},\ \bibinfo {pages} {L161106}
  (\bibinfo {year} {2021})}\BibitemShut {NoStop}%
\bibitem [{\citenamefont {Roccati}(2021)}]{roccati2021non}%
  \BibitemOpen
  \bibfield  {author} {\bibinfo {author} {\bibfnamefont {F.}~\bibnamefont
  {Roccati}},\ }\href@noop {} {\bibfield  {journal} {\bibinfo  {journal} {Phys.
  Rev. A}\ }\textbf {\bibinfo {volume} {104}},\ \bibinfo {pages} {022215}
  (\bibinfo {year} {2021})}\BibitemShut {NoStop}%
\bibitem [{\citenamefont {Silveirinha}(2019)}]{silveirinha2019topological}%
  \BibitemOpen
  \bibfield  {author} {\bibinfo {author} {\bibfnamefont {M.~G.}\ \bibnamefont
  {Silveirinha}},\ }\href@noop {} {\bibfield  {journal} {\bibinfo  {journal}
  {Phys. Rev. B}\ }\textbf {\bibinfo {volume} {99}},\ \bibinfo {pages} {125155}
  (\bibinfo {year} {2019})}\BibitemShut {NoStop}%
\bibitem [{\citenamefont {Pan}\ \emph {et~al.}(2018)\citenamefont {Pan},
  \citenamefont {Zhao}, \citenamefont {Miao}, \citenamefont {Longhi},\ and\
  \citenamefont {Feng}}]{pan2018photonic}%
  \BibitemOpen
  \bibfield  {author} {\bibinfo {author} {\bibfnamefont {M.}~\bibnamefont
  {Pan}}, \bibinfo {author} {\bibfnamefont {H.}~\bibnamefont {Zhao}}, \bibinfo
  {author} {\bibfnamefont {P.}~\bibnamefont {Miao}}, \bibinfo {author}
  {\bibfnamefont {S.}~\bibnamefont {Longhi}},\ and\ \bibinfo {author}
  {\bibfnamefont {L.}~\bibnamefont {Feng}},\ }\href@noop {} {\bibfield
  {journal} {\bibinfo  {journal} {Nat. Commun.}\ }\textbf {\bibinfo {volume}
  {9}},\ \bibinfo {pages} {1} (\bibinfo {year} {2018})}\BibitemShut {NoStop}%
\bibitem [{\citenamefont {Zhou}\ \emph {et~al.}(2021)\citenamefont {Zhou},
  \citenamefont {Li}, \citenamefont {Yi},\ and\ \citenamefont
  {Cui}}]{zhou2021engineering}%
  \BibitemOpen
  \bibfield  {author} {\bibinfo {author} {\bibfnamefont {L.}~\bibnamefont
  {Zhou}}, \bibinfo {author} {\bibfnamefont {H.}~\bibnamefont {Li}}, \bibinfo
  {author} {\bibfnamefont {W.}~\bibnamefont {Yi}},\ and\ \bibinfo {author}
  {\bibfnamefont {X.}~\bibnamefont {Cui}},\ }\href@noop {} {\bibfield
  {journal} {\bibinfo  {journal} {ArXiv preprint arXiv:2111.04196}\ } (\bibinfo
  {year} {2021})}\BibitemShut {NoStop}%
\bibitem [{\citenamefont {Ruschhaupt}\ \emph {et~al.}(2006)\citenamefont
  {Ruschhaupt}, \citenamefont {Muga},\ and\ \citenamefont
  {Raizen}}]{ruschhaupt2006improvement}%
  \BibitemOpen
  \bibfield  {author} {\bibinfo {author} {\bibfnamefont {A.}~\bibnamefont
  {Ruschhaupt}}, \bibinfo {author} {\bibfnamefont {J.}~\bibnamefont {Muga}},\
  and\ \bibinfo {author} {\bibfnamefont {M.}~\bibnamefont {Raizen}},\
  }\href@noop {} {\bibfield  {journal} {\bibinfo  {journal} {J. Phys. B: At.
  Mol. Opt. Phys}\ }\textbf {\bibinfo {volume} {39}},\ \bibinfo {pages} {L133}
  (\bibinfo {year} {2006})}\BibitemShut {NoStop}%
\bibitem [{\citenamefont {Ghatak}\ \emph {et~al.}(2020)\citenamefont {Ghatak},
  \citenamefont {Brandenbourger}, \citenamefont {Van~Wezel},\ and\
  \citenamefont {Coulais}}]{ghatak2020observation}%
  \BibitemOpen
  \bibfield  {author} {\bibinfo {author} {\bibfnamefont {A.}~\bibnamefont
  {Ghatak}}, \bibinfo {author} {\bibfnamefont {M.}~\bibnamefont
  {Brandenbourger}}, \bibinfo {author} {\bibfnamefont {J.}~\bibnamefont
  {Van~Wezel}},\ and\ \bibinfo {author} {\bibfnamefont {C.}~\bibnamefont
  {Coulais}},\ }\href@noop {} {\bibfield  {journal} {\bibinfo  {journal} {Proc.
  Natl. Acad. Sci.}\ }\textbf {\bibinfo {volume} {117}},\ \bibinfo {pages}
  {29561} (\bibinfo {year} {2020})}\BibitemShut {NoStop}%
\bibitem [{\citenamefont {Hou}\ \emph {et~al.}(2020)\citenamefont {Hou},
  \citenamefont {Li}, \citenamefont {Luo}, \citenamefont {Gu},\ and\
  \citenamefont {Zhang}}]{hou2020topological}%
  \BibitemOpen
  \bibfield  {author} {\bibinfo {author} {\bibfnamefont {J.}~\bibnamefont
  {Hou}}, \bibinfo {author} {\bibfnamefont {Z.}~\bibnamefont {Li}}, \bibinfo
  {author} {\bibfnamefont {X.-W.}\ \bibnamefont {Luo}}, \bibinfo {author}
  {\bibfnamefont {Q.}~\bibnamefont {Gu}},\ and\ \bibinfo {author}
  {\bibfnamefont {C.}~\bibnamefont {Zhang}},\ }\href@noop {} {\bibfield
  {journal} {\bibinfo  {journal} {Phys. Rev. Lett.}\ }\textbf {\bibinfo
  {volume} {124}},\ \bibinfo {pages} {073603} (\bibinfo {year}
  {2020})}\BibitemShut {NoStop}%
\bibitem [{\citenamefont {Jin}(2017)}]{jin2017topological}%
  \BibitemOpen
  \bibfield  {author} {\bibinfo {author} {\bibfnamefont {L.}~\bibnamefont
  {Jin}},\ }\href@noop {} {\bibfield  {journal} {\bibinfo  {journal} {Phys.
  Rev. A}\ }\textbf {\bibinfo {volume} {96}},\ \bibinfo {pages} {032103}
  (\bibinfo {year} {2017})}\BibitemShut {NoStop}%
\bibitem [{\citenamefont {Longhi}(2021)}]{longhi2021non}%
  \BibitemOpen
  \bibfield  {author} {\bibinfo {author} {\bibfnamefont {S.}~\bibnamefont
  {Longhi}},\ }\href@noop {} {\bibfield  {journal} {\bibinfo  {journal} {Opt.
  Lett.}\ }\textbf {\bibinfo {volume} {46}},\ \bibinfo {pages} {4470} (\bibinfo
  {year} {2021})}\BibitemShut {NoStop}%
\bibitem [{\citenamefont {Ao}\ \emph {et~al.}(2020)\citenamefont {Ao},
  \citenamefont {Hu}, \citenamefont {You}, \citenamefont {Lu}, \citenamefont
  {Fu}, \citenamefont {Wang},\ and\ \citenamefont {Gong}}]{ao2020topological}%
  \BibitemOpen
  \bibfield  {author} {\bibinfo {author} {\bibfnamefont {Y.}~\bibnamefont
  {Ao}}, \bibinfo {author} {\bibfnamefont {X.}~\bibnamefont {Hu}}, \bibinfo
  {author} {\bibfnamefont {Y.}~\bibnamefont {You}}, \bibinfo {author}
  {\bibfnamefont {C.}~\bibnamefont {Lu}}, \bibinfo {author} {\bibfnamefont
  {Y.}~\bibnamefont {Fu}}, \bibinfo {author} {\bibfnamefont {X.}~\bibnamefont
  {Wang}},\ and\ \bibinfo {author} {\bibfnamefont {Q.}~\bibnamefont {Gong}},\
  }\href@noop {} {\bibfield  {journal} {\bibinfo  {journal} {Phys. Rev. Lett.}\
  }\textbf {\bibinfo {volume} {125}},\ \bibinfo {pages} {013902} (\bibinfo
  {year} {2020})}\BibitemShut {NoStop}%
\bibitem [{\citenamefont {Lin}\ \emph {et~al.}(2021)\citenamefont {Lin},
  \citenamefont {Ke}, \citenamefont {Zhu},\ and\ \citenamefont
  {Li}}]{lin2021square}%
  \BibitemOpen
  \bibfield  {author} {\bibinfo {author} {\bibfnamefont {Z.}~\bibnamefont
  {Lin}}, \bibinfo {author} {\bibfnamefont {S.}~\bibnamefont {Ke}}, \bibinfo
  {author} {\bibfnamefont {X.}~\bibnamefont {Zhu}},\ and\ \bibinfo {author}
  {\bibfnamefont {X.}~\bibnamefont {Li}},\ }\href@noop {} {\bibfield  {journal}
  {\bibinfo  {journal} {Opt. Express}\ }\textbf {\bibinfo {volume} {29}},\
  \bibinfo {pages} {8462} (\bibinfo {year} {2021})}\BibitemShut {NoStop}%
\bibitem [{\citenamefont {Gao}\ \emph {et~al.}(2022)\citenamefont {Gao},
  \citenamefont {Xue}, \citenamefont {Gu}, \citenamefont {Li}, \citenamefont
  {Zhu}, \citenamefont {Su}, \citenamefont {Zhu}, \citenamefont {Zhang},\ and\
  \citenamefont {Chong}}]{gao2022non}%
  \BibitemOpen
  \bibfield  {author} {\bibinfo {author} {\bibfnamefont {H.}~\bibnamefont
  {Gao}}, \bibinfo {author} {\bibfnamefont {H.}~\bibnamefont {Xue}}, \bibinfo
  {author} {\bibfnamefont {Z.}~\bibnamefont {Gu}}, \bibinfo {author}
  {\bibfnamefont {L.}~\bibnamefont {Li}}, \bibinfo {author} {\bibfnamefont
  {W.}~\bibnamefont {Zhu}}, \bibinfo {author} {\bibfnamefont {Z.}~\bibnamefont
  {Su}}, \bibinfo {author} {\bibfnamefont {J.}~\bibnamefont {Zhu}}, \bibinfo
  {author} {\bibfnamefont {B.}~\bibnamefont {Zhang}},\ and\ \bibinfo {author}
  {\bibfnamefont {Y.}~\bibnamefont {Chong}},\ }\href@noop {} {\bibfield
  {journal} {\bibinfo  {journal} {ArXiv preprint arXiv:2205.14824}\ } (\bibinfo
  {year} {2022})}\BibitemShut {NoStop}%
\bibitem [{\citenamefont {Zhang}\ \emph {et~al.}(2023)\citenamefont {Zhang},
  \citenamefont {Zhang}, \citenamefont {Sahin}, \citenamefont {Siu},
  \citenamefont {Rafi-Ul-Islam}, \citenamefont {Kong}, \citenamefont {Shen},
  \citenamefont {Jalil}, \citenamefont {Thomale},\ and\ \citenamefont
  {Lee}}]{zhang2023anomalous}%
  \BibitemOpen
  \bibfield  {author} {\bibinfo {author} {\bibfnamefont {X.}~\bibnamefont
  {Zhang}}, \bibinfo {author} {\bibfnamefont {B.}~\bibnamefont {Zhang}},
  \bibinfo {author} {\bibfnamefont {H.}~\bibnamefont {Sahin}}, \bibinfo
  {author} {\bibfnamefont {Z.~B.}\ \bibnamefont {Siu}}, \bibinfo {author}
  {\bibfnamefont {S.}~\bibnamefont {Rafi-Ul-Islam}}, \bibinfo {author}
  {\bibfnamefont {J.~F.}\ \bibnamefont {Kong}}, \bibinfo {author}
  {\bibfnamefont {B.}~\bibnamefont {Shen}}, \bibinfo {author} {\bibfnamefont
  {M.~B.}\ \bibnamefont {Jalil}}, \bibinfo {author} {\bibfnamefont
  {R.}~\bibnamefont {Thomale}},\ and\ \bibinfo {author} {\bibfnamefont {C.~H.}\
  \bibnamefont {Lee}},\ }\href@noop {} {\bibfield  {journal} {\bibinfo
  {journal} {Commun. Phys.}\ }\textbf {\bibinfo {volume} {6}},\ \bibinfo
  {pages} {151} (\bibinfo {year} {2023})}\BibitemShut {NoStop}%
\bibitem [{\citenamefont {Zou}\ \emph {et~al.}(2021)\citenamefont {Zou},
  \citenamefont {Chen}, \citenamefont {He}, \citenamefont {Bao}, \citenamefont
  {Lee}, \citenamefont {Sun},\ and\ \citenamefont
  {Zhang}}]{zou2021observation}%
  \BibitemOpen
  \bibfield  {author} {\bibinfo {author} {\bibfnamefont {D.}~\bibnamefont
  {Zou}}, \bibinfo {author} {\bibfnamefont {T.}~\bibnamefont {Chen}}, \bibinfo
  {author} {\bibfnamefont {W.}~\bibnamefont {He}}, \bibinfo {author}
  {\bibfnamefont {J.}~\bibnamefont {Bao}}, \bibinfo {author} {\bibfnamefont
  {C.~H.}\ \bibnamefont {Lee}}, \bibinfo {author} {\bibfnamefont
  {H.}~\bibnamefont {Sun}},\ and\ \bibinfo {author} {\bibfnamefont
  {X.}~\bibnamefont {Zhang}},\ }\href@noop {} {\bibfield  {journal} {\bibinfo
  {journal} {Nat. Commun.}\ }\textbf {\bibinfo {volume} {12}},\ \bibinfo
  {pages} {1} (\bibinfo {year} {2021})}\BibitemShut {NoStop}%
\bibitem [{\citenamefont {Rafi-Ul-Islam}\ \emph
  {et~al.}(2024{\natexlab{d}})\citenamefont {Rafi-Ul-Islam}, \citenamefont
  {Siu}, \citenamefont {Sahin},\ and\ \citenamefont {Jalil}}]{rafi2024chiral}%
  \BibitemOpen
  \bibfield  {author} {\bibinfo {author} {\bibfnamefont {S.}~\bibnamefont
  {Rafi-Ul-Islam}}, \bibinfo {author} {\bibfnamefont {Z.~B.}\ \bibnamefont
  {Siu}}, \bibinfo {author} {\bibfnamefont {H.}~\bibnamefont {Sahin}},\ and\
  \bibinfo {author} {\bibfnamefont {M.~B.}\ \bibnamefont {Jalil}},\ }\href@noop
  {} {\bibfield  {journal} {\bibinfo  {journal} {Phys. Rev. B}\ }\textbf
  {\bibinfo {volume} {109}},\ \bibinfo {pages} {085430} (\bibinfo {year}
  {2024}{\natexlab{d}})}\BibitemShut {NoStop}%
\bibitem [{\citenamefont {Rafi-Ul-Islam}\ \emph
  {et~al.}(2020{\natexlab{a}})\citenamefont {Rafi-Ul-Islam}, \citenamefont
  {Siu}, \citenamefont {Sun},\ and\ \citenamefont
  {Jalil}}]{rafi2020realization}%
  \BibitemOpen
  \bibfield  {author} {\bibinfo {author} {\bibfnamefont {S.}~\bibnamefont
  {Rafi-Ul-Islam}}, \bibinfo {author} {\bibfnamefont {Z.~B.}\ \bibnamefont
  {Siu}}, \bibinfo {author} {\bibfnamefont {C.}~\bibnamefont {Sun}},\ and\
  \bibinfo {author} {\bibfnamefont {M.~B.}\ \bibnamefont {Jalil}},\ }\href@noop
  {} {\bibfield  {journal} {\bibinfo  {journal} {New J. Phys.}\ }\textbf
  {\bibinfo {volume} {22}},\ \bibinfo {pages} {023025} (\bibinfo {year}
  {2020}{\natexlab{a}})}\BibitemShut {NoStop}%
\bibitem [{\citenamefont {Sahin}\ \emph {et~al.}(2023)\citenamefont {Sahin},
  \citenamefont {Siu}, \citenamefont {Rafi-Ul-Islam}, \citenamefont {Kong},
  \citenamefont {Jalil},\ and\ \citenamefont {Lee}}]{sahin2023impedance}%
  \BibitemOpen
  \bibfield  {author} {\bibinfo {author} {\bibfnamefont {H.}~\bibnamefont
  {Sahin}}, \bibinfo {author} {\bibfnamefont {Z.~B.}\ \bibnamefont {Siu}},
  \bibinfo {author} {\bibfnamefont {S.}~\bibnamefont {Rafi-Ul-Islam}}, \bibinfo
  {author} {\bibfnamefont {J.~F.}\ \bibnamefont {Kong}}, \bibinfo {author}
  {\bibfnamefont {M.~B.}\ \bibnamefont {Jalil}},\ and\ \bibinfo {author}
  {\bibfnamefont {C.~H.}\ \bibnamefont {Lee}},\ }\href@noop {} {\bibfield
  {journal} {\bibinfo  {journal} {Phys. Rev. B}\ }\textbf {\bibinfo {volume}
  {107}},\ \bibinfo {pages} {245114} (\bibinfo {year} {2023})}\BibitemShut
  {NoStop}%
\bibitem [{\citenamefont {Helbig}\ \emph {et~al.}(2020)\citenamefont {Helbig},
  \citenamefont {Hofmann}, \citenamefont {Imhof}, \citenamefont {Abdelghany},
  \citenamefont {Kiessling}, \citenamefont {Molenkamp}, \citenamefont {Lee},
  \citenamefont {Szameit}, \citenamefont {Greiter},\ and\ \citenamefont
  {Thomale}}]{helbig2020generalized}%
  \BibitemOpen
  \bibfield  {author} {\bibinfo {author} {\bibfnamefont {T.}~\bibnamefont
  {Helbig}}, \bibinfo {author} {\bibfnamefont {T.}~\bibnamefont {Hofmann}},
  \bibinfo {author} {\bibfnamefont {S.}~\bibnamefont {Imhof}}, \bibinfo
  {author} {\bibfnamefont {M.}~\bibnamefont {Abdelghany}}, \bibinfo {author}
  {\bibfnamefont {T.}~\bibnamefont {Kiessling}}, \bibinfo {author}
  {\bibfnamefont {L.}~\bibnamefont {Molenkamp}}, \bibinfo {author}
  {\bibfnamefont {C.}~\bibnamefont {Lee}}, \bibinfo {author} {\bibfnamefont
  {A.}~\bibnamefont {Szameit}}, \bibinfo {author} {\bibfnamefont
  {M.}~\bibnamefont {Greiter}},\ and\ \bibinfo {author} {\bibfnamefont
  {R.}~\bibnamefont {Thomale}},\ }\href@noop {} {\bibfield  {journal} {\bibinfo
   {journal} {Nat. Phys.}\ }\textbf {\bibinfo {volume} {16}},\ \bibinfo {pages}
  {747} (\bibinfo {year} {2020})}\BibitemShut {NoStop}%
\bibitem [{\citenamefont {Rafi-Ul-Islam}\ \emph {et~al.}(2023)\citenamefont
  {Rafi-Ul-Islam}, \citenamefont {Siu}, \citenamefont {Sahin},\ and\
  \citenamefont {Jalil}}]{rafi2023valley}%
  \BibitemOpen
  \bibfield  {author} {\bibinfo {author} {\bibfnamefont {S.}~\bibnamefont
  {Rafi-Ul-Islam}}, \bibinfo {author} {\bibfnamefont {Z.~B.}\ \bibnamefont
  {Siu}}, \bibinfo {author} {\bibfnamefont {H.}~\bibnamefont {Sahin}},\ and\
  \bibinfo {author} {\bibfnamefont {M.~B.}\ \bibnamefont {Jalil}},\ }\href@noop
  {} {\bibfield  {journal} {\bibinfo  {journal} {Phys. Rev. Res.}\ }\textbf
  {\bibinfo {volume} {5}},\ \bibinfo {pages} {013107} (\bibinfo {year}
  {2023})}\BibitemShut {NoStop}%
\bibitem [{\citenamefont {Zhang}\ \emph {et~al.}(2022)\citenamefont {Zhang},
  \citenamefont {Wu}, \citenamefont {Liu}, \citenamefont {Xiao},\ and\
  \citenamefont {Liu}}]{zhang2022complex}%
  \BibitemOpen
  \bibfield  {author} {\bibinfo {author} {\bibfnamefont {R.-L.}\ \bibnamefont
  {Zhang}}, \bibinfo {author} {\bibfnamefont {Q.-P.}\ \bibnamefont {Wu}},
  \bibinfo {author} {\bibfnamefont {M.-R.}\ \bibnamefont {Liu}}, \bibinfo
  {author} {\bibfnamefont {X.-B.}\ \bibnamefont {Xiao}},\ and\ \bibinfo
  {author} {\bibfnamefont {Z.-F.}\ \bibnamefont {Liu}},\ }\href@noop {}
  {\bibfield  {journal} {\bibinfo  {journal} {Annalen der Physik}\ ,\ \bibinfo
  {pages} {2100497}} (\bibinfo {year} {2022})}\BibitemShut {NoStop}%
\bibitem [{\citenamefont {Hofmann}\ \emph {et~al.}(2020)\citenamefont
  {Hofmann}, \citenamefont {Helbig}, \citenamefont {Schindler}, \citenamefont
  {Salgo}, \citenamefont {Brzezi{\'n}ska}, \citenamefont {Greiter},
  \citenamefont {Kiessling}, \citenamefont {Wolf}, \citenamefont {Vollhardt},
  \citenamefont {Kaba{\v{s}}i} \emph {et~al.}}]{hofmann2020reciprocal}%
  \BibitemOpen
  \bibfield  {author} {\bibinfo {author} {\bibfnamefont {T.}~\bibnamefont
  {Hofmann}}, \bibinfo {author} {\bibfnamefont {T.}~\bibnamefont {Helbig}},
  \bibinfo {author} {\bibfnamefont {F.}~\bibnamefont {Schindler}}, \bibinfo
  {author} {\bibfnamefont {N.}~\bibnamefont {Salgo}}, \bibinfo {author}
  {\bibfnamefont {M.}~\bibnamefont {Brzezi{\'n}ska}}, \bibinfo {author}
  {\bibfnamefont {M.}~\bibnamefont {Greiter}}, \bibinfo {author} {\bibfnamefont
  {T.}~\bibnamefont {Kiessling}}, \bibinfo {author} {\bibfnamefont
  {D.}~\bibnamefont {Wolf}}, \bibinfo {author} {\bibfnamefont {A.}~\bibnamefont
  {Vollhardt}}, \bibinfo {author} {\bibfnamefont {A.}~\bibnamefont
  {Kaba{\v{s}}i}}, \emph {et~al.},\ }\href@noop {} {\bibfield  {journal}
  {\bibinfo  {journal} {Phys. Rev. Res.}\ }\textbf {\bibinfo {volume} {2}},\
  \bibinfo {pages} {023265} (\bibinfo {year} {2020})}\BibitemShut {NoStop}%
\bibitem [{\citenamefont {Shang}\ \emph {et~al.}(2022)\citenamefont {Shang},
  \citenamefont {Liu}, \citenamefont {Shao}, \citenamefont {Han}, \citenamefont
  {Zang}, \citenamefont {Zhang}, \citenamefont {Salama}, \citenamefont {Gao},
  \citenamefont {Lee}, \citenamefont {Thomale} \emph
  {et~al.}}]{shang2022experimental}%
  \BibitemOpen
  \bibfield  {author} {\bibinfo {author} {\bibfnamefont {C.}~\bibnamefont
  {Shang}}, \bibinfo {author} {\bibfnamefont {S.}~\bibnamefont {Liu}}, \bibinfo
  {author} {\bibfnamefont {R.}~\bibnamefont {Shao}}, \bibinfo {author}
  {\bibfnamefont {P.}~\bibnamefont {Han}}, \bibinfo {author} {\bibfnamefont
  {X.}~\bibnamefont {Zang}}, \bibinfo {author} {\bibfnamefont {X.}~\bibnamefont
  {Zhang}}, \bibinfo {author} {\bibfnamefont {K.~N.}\ \bibnamefont {Salama}},
  \bibinfo {author} {\bibfnamefont {W.}~\bibnamefont {Gao}}, \bibinfo {author}
  {\bibfnamefont {C.~H.}\ \bibnamefont {Lee}}, \bibinfo {author} {\bibfnamefont
  {R.}~\bibnamefont {Thomale}}, \emph {et~al.},\ }\href@noop {} {\bibfield
  {journal} {\bibinfo  {journal} {ArXiv preprint arXiv:2203.00484}\ } (\bibinfo
  {year} {2022})}\BibitemShut {NoStop}%
\bibitem [{\citenamefont {Liu}\ \emph {et~al.}(2020)\citenamefont {Liu},
  \citenamefont {Ma}, \citenamefont {Yang}, \citenamefont {Zhang},
  \citenamefont {Gao}, \citenamefont {Xiang}, \citenamefont {Cui},\ and\
  \citenamefont {Zhang}}]{liu2020gain}%
  \BibitemOpen
  \bibfield  {author} {\bibinfo {author} {\bibfnamefont {S.}~\bibnamefont
  {Liu}}, \bibinfo {author} {\bibfnamefont {S.}~\bibnamefont {Ma}}, \bibinfo
  {author} {\bibfnamefont {C.}~\bibnamefont {Yang}}, \bibinfo {author}
  {\bibfnamefont {L.}~\bibnamefont {Zhang}}, \bibinfo {author} {\bibfnamefont
  {W.}~\bibnamefont {Gao}}, \bibinfo {author} {\bibfnamefont {Y.~J.}\
  \bibnamefont {Xiang}}, \bibinfo {author} {\bibfnamefont {T.~J.}\ \bibnamefont
  {Cui}},\ and\ \bibinfo {author} {\bibfnamefont {S.}~\bibnamefont {Zhang}},\
  }\href@noop {} {\bibfield  {journal} {\bibinfo  {journal} {Phys. Rev. Appl.}\
  }\textbf {\bibinfo {volume} {13}},\ \bibinfo {pages} {014047} (\bibinfo
  {year} {2020})}\BibitemShut {NoStop}%
\bibitem [{\citenamefont {Kotwal}\ \emph {et~al.}(2021)\citenamefont {Kotwal},
  \citenamefont {Moseley}, \citenamefont {Stegmaier}, \citenamefont {Imhof},
  \citenamefont {Brand}, \citenamefont {Kie{\ss}ling}, \citenamefont {Thomale},
  \citenamefont {Ronellenfitsch},\ and\ \citenamefont
  {Dunkel}}]{kotwal2021active}%
  \BibitemOpen
  \bibfield  {author} {\bibinfo {author} {\bibfnamefont {T.}~\bibnamefont
  {Kotwal}}, \bibinfo {author} {\bibfnamefont {F.}~\bibnamefont {Moseley}},
  \bibinfo {author} {\bibfnamefont {A.}~\bibnamefont {Stegmaier}}, \bibinfo
  {author} {\bibfnamefont {S.}~\bibnamefont {Imhof}}, \bibinfo {author}
  {\bibfnamefont {H.}~\bibnamefont {Brand}}, \bibinfo {author} {\bibfnamefont
  {T.}~\bibnamefont {Kie{\ss}ling}}, \bibinfo {author} {\bibfnamefont
  {R.}~\bibnamefont {Thomale}}, \bibinfo {author} {\bibfnamefont
  {H.}~\bibnamefont {Ronellenfitsch}},\ and\ \bibinfo {author} {\bibfnamefont
  {J.}~\bibnamefont {Dunkel}},\ }\href@noop {} {\bibfield  {journal} {\bibinfo
  {journal} {Proc. Natl. Acad. Sci.}\ }\textbf {\bibinfo {volume} {118}},\
  \bibinfo {pages} {e2106411118} (\bibinfo {year} {2021})}\BibitemShut
  {NoStop}%
\bibitem [{\citenamefont {Lee}\ \emph {et~al.}(2018)\citenamefont {Lee},
  \citenamefont {Imhof}, \citenamefont {Berger}, \citenamefont {Bayer},
  \citenamefont {Brehm}, \citenamefont {Molenkamp}, \citenamefont {Kiessling},\
  and\ \citenamefont {Thomale}}]{lee2018topolectrical}%
  \BibitemOpen
  \bibfield  {author} {\bibinfo {author} {\bibfnamefont {C.~H.}\ \bibnamefont
  {Lee}}, \bibinfo {author} {\bibfnamefont {S.}~\bibnamefont {Imhof}}, \bibinfo
  {author} {\bibfnamefont {C.}~\bibnamefont {Berger}}, \bibinfo {author}
  {\bibfnamefont {F.}~\bibnamefont {Bayer}}, \bibinfo {author} {\bibfnamefont
  {J.}~\bibnamefont {Brehm}}, \bibinfo {author} {\bibfnamefont {L.~W.}\
  \bibnamefont {Molenkamp}}, \bibinfo {author} {\bibfnamefont {T.}~\bibnamefont
  {Kiessling}},\ and\ \bibinfo {author} {\bibfnamefont {R.}~\bibnamefont
  {Thomale}},\ }\href@noop {} {\bibfield  {journal} {\bibinfo  {journal}
  {Commun. Phys.}\ }\textbf {\bibinfo {volume} {1}},\ \bibinfo {pages} {1}
  (\bibinfo {year} {2018})}\BibitemShut {NoStop}%
\bibitem [{\citenamefont {Rafi-Ul-Islam}\ \emph
  {et~al.}(2020{\natexlab{b}})\citenamefont {Rafi-Ul-Islam}, \citenamefont
  {Bin~Siu},\ and\ \citenamefont {Jalil}}]{rafi2020topoelectrical}%
  \BibitemOpen
  \bibfield  {author} {\bibinfo {author} {\bibfnamefont {S.}~\bibnamefont
  {Rafi-Ul-Islam}}, \bibinfo {author} {\bibfnamefont {Z.}~\bibnamefont
  {Bin~Siu}},\ and\ \bibinfo {author} {\bibfnamefont {M.}~\bibnamefont
  {Jalil}},\ }\href@noop {} {\bibfield  {journal} {\bibinfo  {journal} {Commun.
  Phys.}\ }\textbf {\bibinfo {volume} {3}},\ \bibinfo {pages} {1} (\bibinfo
  {year} {2020}{\natexlab{b}})}\BibitemShut {NoStop}%
\bibitem [{\citenamefont {Dong}\ \emph {et~al.}(2021)\citenamefont {Dong},
  \citenamefont {Juri{\v{c}}i{\'c}},\ and\ \citenamefont
  {Roy}}]{dong2021topolectric}%
  \BibitemOpen
  \bibfield  {author} {\bibinfo {author} {\bibfnamefont {J.}~\bibnamefont
  {Dong}}, \bibinfo {author} {\bibfnamefont {V.}~\bibnamefont
  {Juri{\v{c}}i{\'c}}},\ and\ \bibinfo {author} {\bibfnamefont
  {B.}~\bibnamefont {Roy}},\ }\href@noop {} {\bibfield  {journal} {\bibinfo
  {journal} {Phys. Rev. Res.}\ }\textbf {\bibinfo {volume} {3}},\ \bibinfo
  {pages} {023056} (\bibinfo {year} {2021})}\BibitemShut {NoStop}%
\bibitem [{\citenamefont {Rafi-Ul-Islam}\ \emph
  {et~al.}(2020{\natexlab{c}})\citenamefont {Rafi-Ul-Islam}, \citenamefont
  {Siu},\ and\ \citenamefont {Jalil}}]{rafi2020anti}%
  \BibitemOpen
  \bibfield  {author} {\bibinfo {author} {\bibfnamefont {S.}~\bibnamefont
  {Rafi-Ul-Islam}}, \bibinfo {author} {\bibfnamefont {Z.~B.}\ \bibnamefont
  {Siu}},\ and\ \bibinfo {author} {\bibfnamefont {M.~B.}\ \bibnamefont
  {Jalil}},\ }\href@noop {} {\bibfield  {journal} {\bibinfo  {journal} {Appl.
  Phys. Lett.}\ }\textbf {\bibinfo {volume} {116}},\ \bibinfo {pages} {111904}
  (\bibinfo {year} {2020}{\natexlab{c}})}\BibitemShut {NoStop}%
\bibitem [{\citenamefont {Zhang}\ \emph {et~al.}(2020)\citenamefont {Zhang},
  \citenamefont {Zou}, \citenamefont {Bao}, \citenamefont {He}, \citenamefont
  {Pei}, \citenamefont {Sun},\ and\ \citenamefont
  {Zhang}}]{zhang2020topolectrical}%
  \BibitemOpen
  \bibfield  {author} {\bibinfo {author} {\bibfnamefont {W.}~\bibnamefont
  {Zhang}}, \bibinfo {author} {\bibfnamefont {D.}~\bibnamefont {Zou}}, \bibinfo
  {author} {\bibfnamefont {J.}~\bibnamefont {Bao}}, \bibinfo {author}
  {\bibfnamefont {W.}~\bibnamefont {He}}, \bibinfo {author} {\bibfnamefont
  {Q.}~\bibnamefont {Pei}}, \bibinfo {author} {\bibfnamefont {H.}~\bibnamefont
  {Sun}},\ and\ \bibinfo {author} {\bibfnamefont {X.}~\bibnamefont {Zhang}},\
  }\href@noop {} {\bibfield  {journal} {\bibinfo  {journal} {Phys. Rev. B}\
  }\textbf {\bibinfo {volume} {102}},\ \bibinfo {pages} {100102} (\bibinfo
  {year} {2020})}\BibitemShut {NoStop}%
\bibitem [{\citenamefont {Kunst}\ \emph {et~al.}(2018)\citenamefont {Kunst},
  \citenamefont {Edvardsson}, \citenamefont {Budich},\ and\ \citenamefont
  {Bergholtz}}]{kunst2018biorthogonal}%
  \BibitemOpen
  \bibfield  {author} {\bibinfo {author} {\bibfnamefont {F.~K.}\ \bibnamefont
  {Kunst}}, \bibinfo {author} {\bibfnamefont {E.}~\bibnamefont {Edvardsson}},
  \bibinfo {author} {\bibfnamefont {J.~C.}\ \bibnamefont {Budich}},\ and\
  \bibinfo {author} {\bibfnamefont {E.~J.}\ \bibnamefont {Bergholtz}},\
  }\href@noop {} {\bibfield  {journal} {\bibinfo  {journal} {Phys. Rev. Lett.}\
  }\textbf {\bibinfo {volume} {121}},\ \bibinfo {pages} {026808} (\bibinfo
  {year} {2018})}\BibitemShut {NoStop}%
\bibitem [{\citenamefont {Yang}\ \emph {et~al.}(2020)\citenamefont {Yang},
  \citenamefont {Zhang}, \citenamefont {Fang},\ and\ \citenamefont
  {Hu}}]{yang2020non}%
  \BibitemOpen
  \bibfield  {author} {\bibinfo {author} {\bibfnamefont {Z.}~\bibnamefont
  {Yang}}, \bibinfo {author} {\bibfnamefont {K.}~\bibnamefont {Zhang}},
  \bibinfo {author} {\bibfnamefont {C.}~\bibnamefont {Fang}},\ and\ \bibinfo
  {author} {\bibfnamefont {J.}~\bibnamefont {Hu}},\ }\href@noop {} {\bibfield
  {journal} {\bibinfo  {journal} {Phys. Rev. Lett.}\ }\textbf {\bibinfo
  {volume} {125}},\ \bibinfo {pages} {226402} (\bibinfo {year}
  {2020})}\BibitemShut {NoStop}%
\end{thebibliography}


\end{document}